\documentclass[oneside,frenchb,english]{aa}
\usepackage[latin1]{inputenc}
\usepackage{array}
\usepackage{float}
\usepackage{color}

\makeatletter

\usepackage{graphicx}
\usepackage{txfonts}
\usepackage{babel}
\makeatother
\begin{document}

\title{Analysis of the circumstellar environment of the B{[}e{]} star \\HD
45677 (FS CMa)\thanks {Based on observations obtained at the Haute Provence Observatory. Table.12 is available in electronic at the CDS via http://vizier.u-strasbg.fr (130.79.128.13) or via http://cdsweb.u-strasbg.fr/cats/Cats.htx (130.79.128.40)}}

\author{G. Muratorio \inst{1}, C. Rossi \inst{2}, M. Friedjung \inst{3}\authorrunning {Muratorio \& al.}}

\institute{Laboratoire d'Astrophysique de Marseille, 2 Place le Verrier, FR-13248
Marseille Cedex 04, France\\
 \email{gerard.muratorio@oamp.fr}\\
 \and
Dipartimento di Fisica , Universit\`a {}``La Sapienza'', Piazzzale
A. Moro 2 00185, Roma, Italy\\
 \email{corinne.rossi@roma1.infn.it}\\
 \and
Institut d'Astrophysique, 98 bis Boulevard Arago, FR-75014 Paris,
France\\
 \email{fried@iap.fr}}

\offprints{G. Muratorio  \email{gerard.muratorio@oamp.fr}}

\abstract{We studied the circumstellar environment of the  B{[}e{]} star HD 45677 through the analysis of the emission lines from ionized metals.
We used the statistical approach of the self absorption curve method (SAC) to derive physical parameters of the line emitting region.
The Fe II and Cr II double-peaked emission line structure is explained by the presence of a thin absorption component red shifted by $\sim$ 3
$\mathrm{km\, s^{-1}}$. This absorption component can be interpreted geometricaly as being due to infalling material perpendicularly to the disk seen
nearly pole-on, as indicated by the emission line structure. The Cr II and Fe II emission lines have a complex structure with two (narrow and broad) components,
of 45 and 180 $\mathrm{km\, s^{-1}}$ FWHM for the permitted lines and 25 and 100 $\mathrm{km\, s^{-1}}$ FWHM for the forbidden ones, respectively.
From our best data set of 1999, we obtained a Boltzmann-type population law whose excitation temperature
is $3\,900_{-600}^{+900}\,\mathrm{K}$ and $3\,150_{-300}^{+350}\,\mathrm{K}$ for the narrow component lower and upper levels respectively.
We obtained an excitation temperature of $3\,400_{-300}^{+350}\,\mathrm{K}$ for the broad component upper levels. The forbidden lines are found to be
formed in the outer regions with higher excitation temperatures of $10\, 500 \pm 1\, 000 K$ and $8\, 000 \pm 1\, 500 K$ for the narrow and broad components
respectively in 1999. Our results are consistent with line formation in a rotating disk, around a young star. In the framework of a very simplified geometrical
model, we argue that the narrow components are principaly emitted by an optically thin disk seen nearly pole-on, in a region whose minimum radius is estimated
to be $4\,10^{12}cm$, while the broad ones are formed in a disk-linked wind.}

\maketitle


\section{Introduction}

HD 45677 (FS CMa) is generally recognized as an early type star, usually
classified as B{[}e{]}, or more precisely as showing the B{[}e{]}
phenomenon, even if it is difficult to determine its exact evolutionary
stage (de Winter et al. \cite{dWvdA97}; Lamers et al.\cite{LZWHZ98}).

The presence of a disk, and even more of a bipolar environment, was
inferred by polarization measurements (Coyne \& Vrba \cite{CV76};
Schulte-Ladbeck et al. \cite{SLSNCABBCMMTW93}) and by UV spectroscopic
measurements of accreting gas (Grady et al. \cite{GBS93}). These
last authors agreed on the presence, around the star, of an actively
accreting circumstellar, protoplanetary disk, and the presence of
low-velocity absorption profiles in Fe {\scriptsize II} lines, among
other species, while optical spectroscopic measurements of strong
absorption cores of a few lines are commented on by de Winter et al.
(\cite{dWvdA97}).

In the optical HD 45677 is characterised by a rich emission line spectrum,
including the Balmer lines of hydrogen, and both permitted and forbidden
lines of neutral and ionized metals (Mg {\scriptsize I}, Mg {\scriptsize II},
Na {\scriptsize I}, Cr {\scriptsize II}, Si {\scriptsize II}, Mn {\scriptsize II},
Ti{\scriptsize ~II}, {[}Ni{\scriptsize ~II}{]}, {[}S{\scriptsize ~II{]}}
and especially Fe {\scriptsize II} and {[}Fe {\scriptsize II}{]}). 

Israelian et al. (1996) found a surface gravity of $\log\,\mathrm{g}=3.85\pm0.05$
(luminosity class V) from the photospheric wings of H$\gamma$ and
H$\delta$, while examination of the He {\scriptsize I} absorption
lines suggested a spectral class of B2~($\mathrm{\mathrm{T_{eff}=}22\,000\pm1\,500\, K}$).
The effective temperature was less certain than the surface gravity,
due to a circumstellar contribution to the He {\scriptsize I} lines.
Israelian et al. (\cite{Ia96}) also found that the photospheric Si
{\scriptsize II} 4128 and 4130 $\textrm{Å}$~lines were not blended,
and could be only fitted by models with a projected rotational velocity
$v\,\sin i$ of less than 70 $\mathrm{km\, s^{-1}}$.

Cidale et al. (\cite{CZT01}) used the method of Barbier, Chalonge
\& Divan, based on the magnitude and position of the Balmer \textit{photospheric}
(not circumstellar) discontinuity. They obtained a spectral type of
B2IV-V, corresponding to $\mathrm{\mathrm{T_{eff}=}21\,500\pm300\, K}$
and $\log\, g=3.89$, in agreement with the previous determinations.

De Winter \& van den Ancker (\cite{dWvdA97}) studied the behaviour
of HD 45677 during the last 25 years, and concluded that the large
photometric variations of the star are best explained by a variable
obscuration from circumstellar grains. They also derived a stellar
instrinsic visual--ultraviolet energy distribution close to that of
a B2 V star, with an infrared excess due to the dust emission. The
recent spectral variations were discussed by Israelian et al. (\cite{Ia96})
and Israelian \& Musaev (\cite{IM97}).

In the present study, we analyse the emission line spectrum in order
to get an insight into the physical processes of line excitation and
geometrical conditions of their forming regions. It is clearly impossible
at the present stage to calculate a detailed synthetic spectrum for
this badly known object, so in order to analyse the spectrum, we use
the statistical approach offered by the Self-Absorption Curve method
(Friedjung \& Muratorio \cite{FM87}; Muratorio \& Friedjung \cite{MF88};
Baratta et al. \cite{Ba98}; Kotnik-Karuza et al. \cite{KFS02}).

\section{Observations and Reductions}

The spectra of HD 45677 were obtained in various contiguous domains
in the course of different observing runs at the Haute Provence Observatory
(OHP) by means of the Aurelie spectrograph attached to the Coudé focus
of the 1.52 m telescope.

Until 1999 Aurelie used as receiver a Thomson TH 7832 photocell {}``Double
Barette'', and since 2000 a new CCD/EEV which improved the instrument
performances (see http://.www.obs-hp.fr/www/guide/aurelie/annonce.html).

We obtained many spectra in selected regions; A higher resolution
was used in 1999 and 2001-2002 in order to characterise better the
profile of the metallic spectral lines. A detailed log of the observations
is given in Table.11 (online material).

The 1995 observations were used in 1996 for the analysis of the He
{\scriptsize I} lines 4387, 4471, 5876 and 4438 $\textrm{Å}$ (Israelian
et al. \cite{Ia96}).

The spectra were reduced at the Marseille Observatory using procedures
runing in the MIDAS environment.

\begin{figure*}
\includegraphics[%
  bb=55bp 100bp 532bp 780bp,
  clip,
  width=8cm,
  height=18cm,
  angle=270]{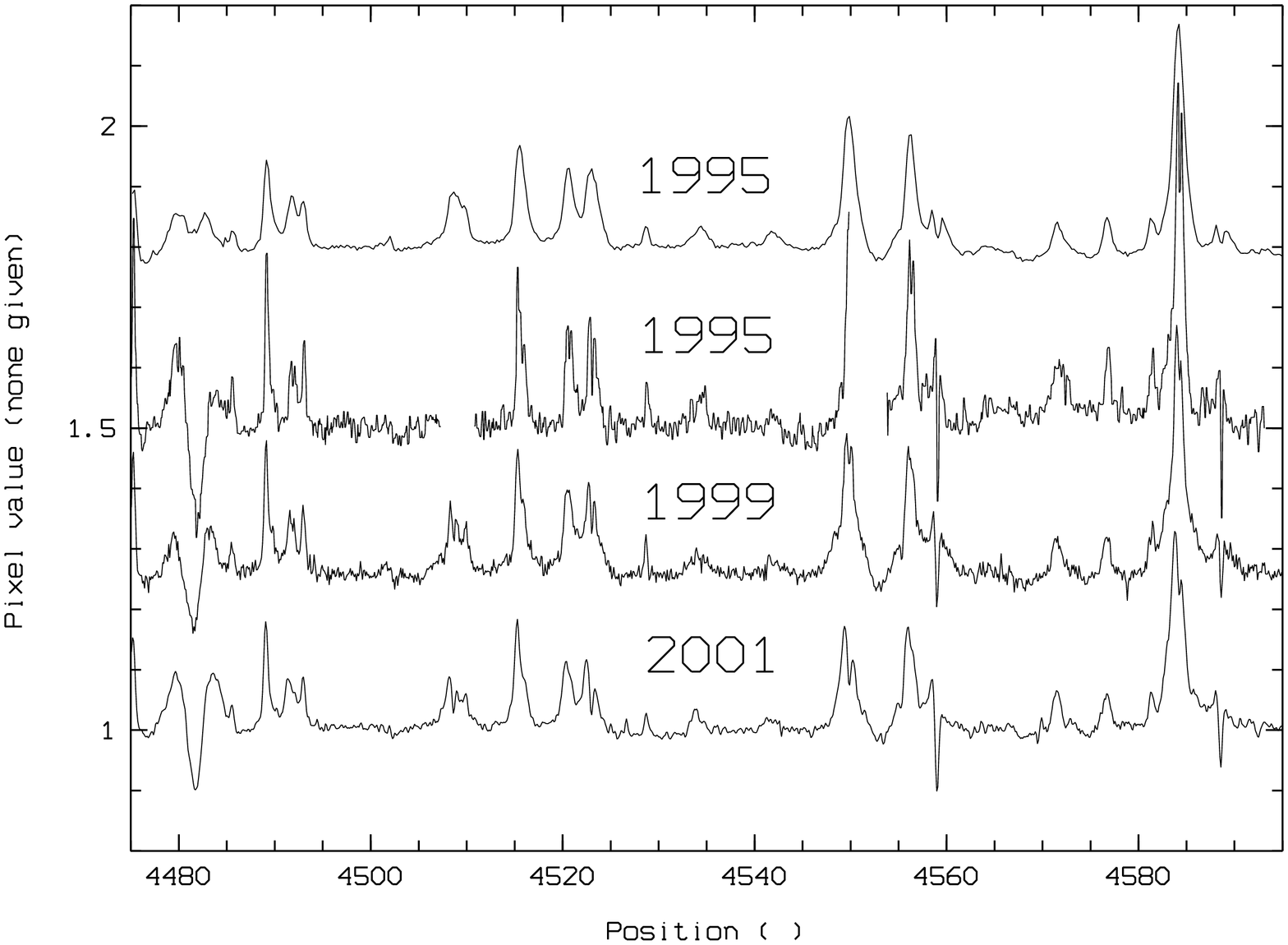}

\caption{The spectrum of HD 45677 during 1995-2001. The spectrograms, normalized
to the continuum and vertically shifted, are, from top to bottom,
of 14/01/1995 (R=7000), 19/10/1995 (R=36000), 2/2/1999 (R=15000) and
5/12/2001 (R=15000). All the spectra were obtained at OHP with AURELIE,
except the October 1995 one which is from the Russian Special Astrophysical
Observatory. On the higher resolution spectrograms the deep Cr {\scriptsize II}
absorptions at 4558 and 4588 $\textrm{Å}$~show up clearly.\label{Parts_spec}}
\end{figure*}

After the files were calibrated in wavelength, and reduced to the
heliocentric standard of rest, we identified the spectral features
present in all our spectra. No large variations are present in the
metallic emission lines within our observing time target (see Fig.~\ref{Parts_spec}).
Higher resolution spectra of HD 45677 were obtained from October 14,
19 and 20 in 1995 by Musaev with the Coudé Echelle Spectrometer at
the 1-m telescope of the Special Astrophysical Observatory of the
Russian Academy of Sciences (Israelian \& Musaev \cite{IM97}), and
kindly provided for us by Garik Israelian.

For the purpose of our analysis of the emission lines we need the
absolute values of the fluxes. To do this, we have normalized the
individual spectra to the local continuum level and, following de
Winter et al. (\cite{dWvdA97}), we multiplied the line fluxes by
the energy distribution of the Kurucz (\cite{K79}, \cite{K93}) model
of $\mathrm{T_{eff}=22\,500 K}$
and $\mathrm{\log\, g=4.0}$ with solar abundance, normalized
to the visual magnitude of the star, V$_{\textrm{0}}$=6.6. The results
are not dependent on the model choosen in the limit of the precision
of the determination of the star's type by various authors (see Introduction).
\textbf{}We adopted the same continuum distribution for the three
studied epochs.

\subsection{Line identification}

Extensive line identification lists of HD 45677 are available from
the literature (Swings \cite{S73}; Andrillat et al. \cite{Aa97};
de Winter et al. \cite{dWvdA97}). However, they generally refer either
to lower resolution spectra and/or to limited spectral ranges. We
therefore prepared a new line list, based on our observations, which
is presented in Table.12 (online material). Emission and
absorption components identified in different species are summarized
in Table.13 (online material).

\subsection{Line components \label{sub:Line-components}}

Figure~\ref{Parts_spec} shows that the lines of the most numerous
species, namely Cr {\scriptsize II} , Fe {\scriptsize II} and {[}Fe
{\scriptsize II} {]}, did not significantly vary during 1995-2002.
This will be best illustrated at the end of this section.

\begin{figure}[h]
\includegraphics[%
  bb=55bp 115bp 532bp 780bp,
  clip,
  width=5cm,
  height=9cm,
  angle=270]{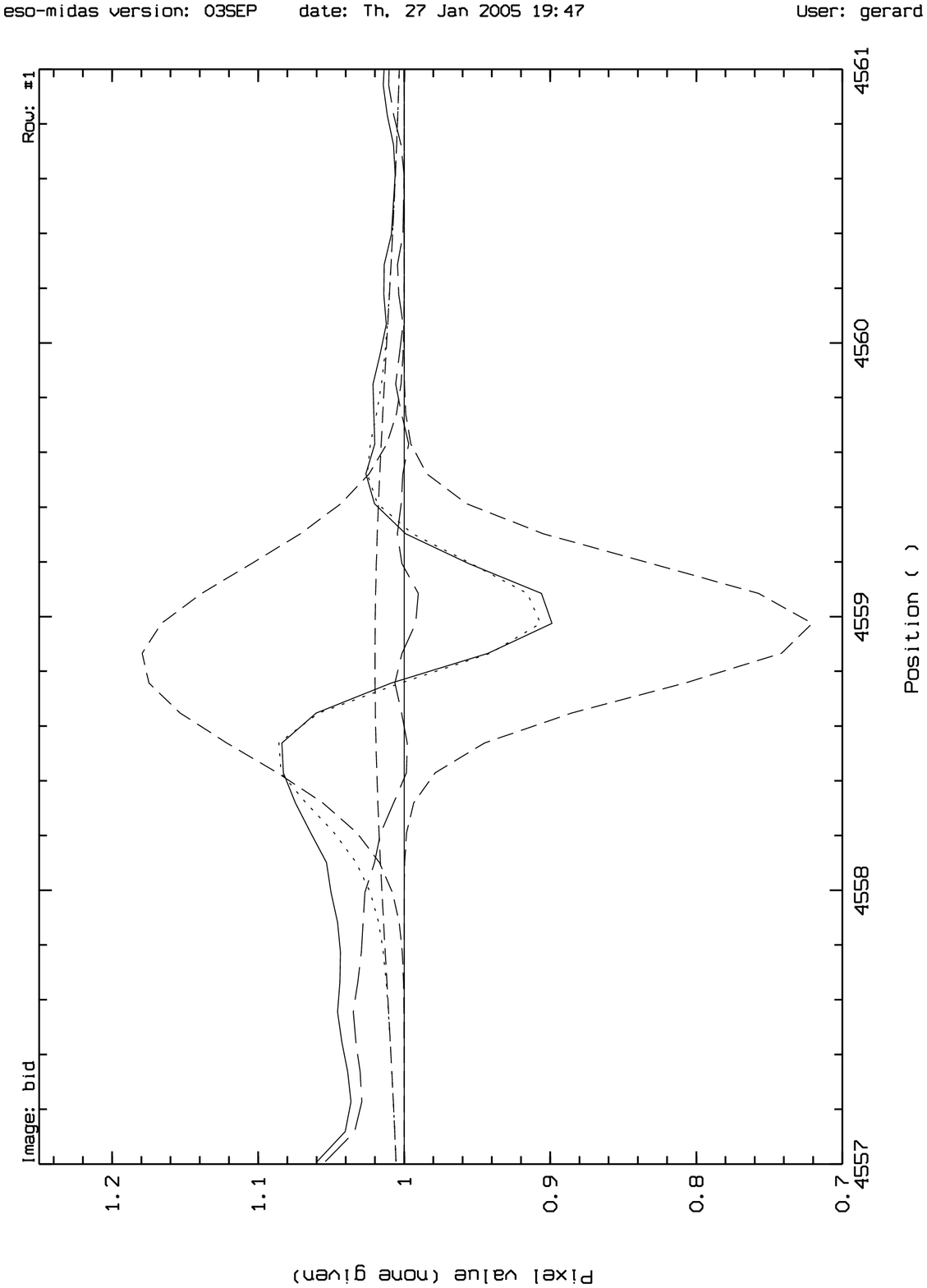}

\includegraphics[%
  bb=55bp 115bp 532bp 780bp,
  clip,
  width=5cm,
  height=9cm,
  angle=270]{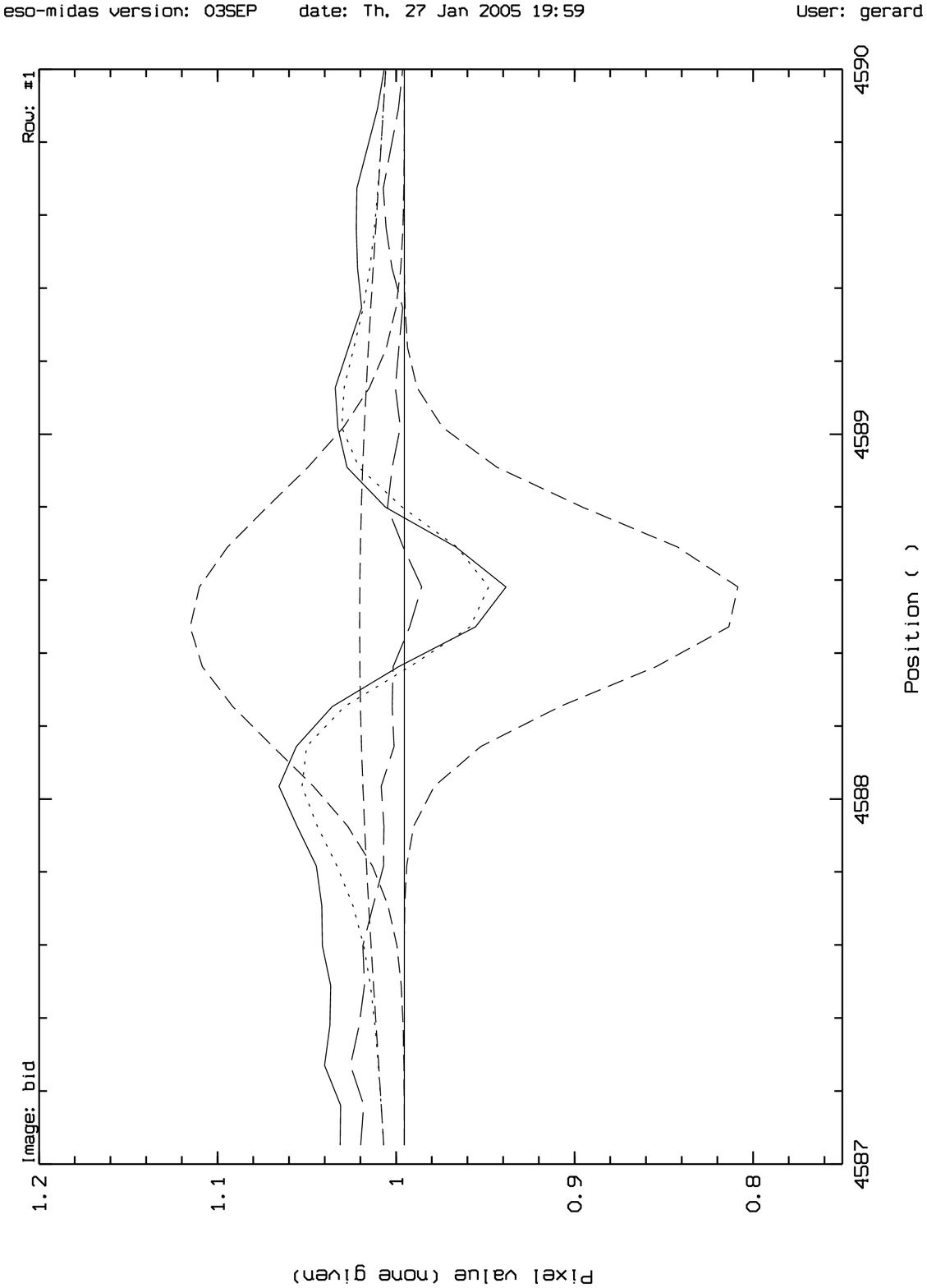}

\caption{The two most intense lines of multiplet 44 of Cr {\scriptsize II}
observed the 5/12/2001, showing the central absorption well below
the continuum and the three Gaussian components' fit . Each line profile
(solid line) is fitted (dotted line) by three Gaussian components
(dashed line). The residual (long dashed line) appears around the
continuum (straight line). Notice the rising of the residual towards
shorter wavelengths which is in both cases due to the presence of
a strong Fe {\scriptsize II} line. \label{Fig_CrII_prof}}
\end{figure}

\begin{figure}[h]
\includegraphics[%
  bb=55bp 115bp 532bp 780bp,
  clip,
  width=7cm,
  height=9cm,
  angle=270]{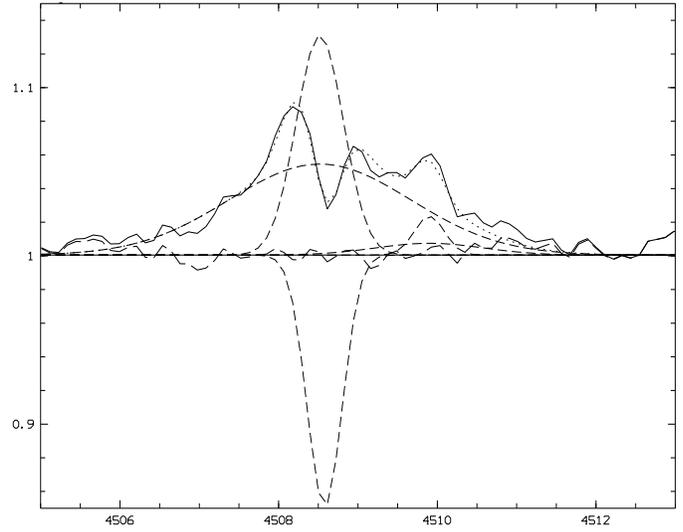}

\caption{Profiles and various components of the blend 4508.288 $\textrm{Å}$
(Fe {\scriptsize II} multiplet 38), 4509.610 $\textrm{Å}$ ({[}Fe
{\scriptsize II}{]} multiplet 6F) on the 2/2/1999 (see Fig.~\ref{Fig_CrII_prof}
for the explanation of the different curves). \textbf{}The different
component FWHM are given in the text.\label{Prof_blend_4508}}
\end{figure}

\begin{figure}[h]
\includegraphics[%
  bb=55bp 115bp 532bp 780bp,
  clip,
  width=7cm,
  height=9cm,
  angle=270]{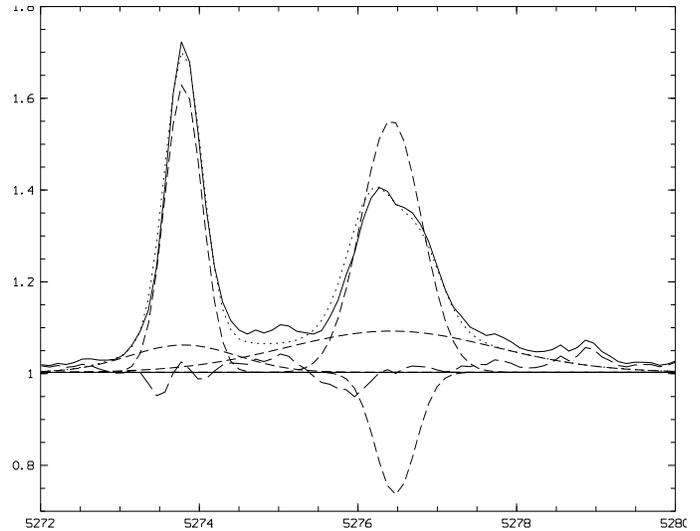}

\caption{Profiles and various components of 5273.38~$\textrm{Å}$ ({[}Fe
{\scriptsize II}{]} multiplet 18F), 5276.002~$\textrm{Å}$ (Fe {\scriptsize II}
multiplet 49), and 5278.374~$\textrm{Å}$ ({[}Fe {\scriptsize II}{]}
multiplet 35F) on the 3/2/1999 (see Fig.~\ref{Fig_CrII_prof} for
the explanation of the different curves).The different component FWHM
are given in the text.\label{Prof_blend_5273_5276}}
\end{figure}

As shown in Figure \ref{Parts_spec}, and, more in detail, in Figures
\ref{Fig_CrII_prof}, \ref{Prof_blend_4508}, \ref{Prof_blend_5273_5276}
and 17 (online material), in the higher resolution spectra
of HD 45677, the strongest permitted lines present a double peaked
emission. This feature has been in previous papers ascribed to emission
from a rotating equatorial disk (de Winter et al. \cite{dWvdA97}).
But they can be as well attributed to absorption in the emitting medium
or in a circumstellar shell according to Grady et al. (\cite{GBS93})
from their ultraviolet observations.

Israelian and Musaev (\cite{IM97}) discussed the narrow absorption
components observed in Mg {\scriptsize II} $\lambda\:4481\,\textrm{Å}$
and in the hydrogen lines in the framework of both infalling material
with a patchy structure, and outfall-wind. They also noticed the strong
variations of the Mg {\scriptsize II} line from pure absorption to
P Cygni profile on timescales of a few days.

Actually, in the higher resolution spectra the central dips of the
strong Mg {\scriptsize II} $\lambda\:4481\,\textrm{Å}$ and Cr {\scriptsize II}
$\lambda\:4558.6\,\textrm{Å}$ and $\lambda\:4588.2\,\textrm{Å}$
go well below the continuum level (Figs.~\ref{Parts_spec} and \ref{Fig_CrII_prof}).

In our lower resolution 1995 spectra (top), the absorption is probably
smoothed out, but is nevertheless present, as confirmed by the fitting
procedure discussed below. The other emission lines are either sligthly
flat--topped or present a dip at the red side of their maximum, again
suggesting the presence of an unresolved central (or slightly red--shifted)
absorption.

Indeed, the line shape is that typical of an emission line with a
central absorption, rather than that of a rotating disk seen close
to edge-on and we shall see in Section~\ref{ssec:Model} that these
profiles can be interpreted as originating in a nearly face-on seen
disk.

At the same time, the strongest emission lines clearly present extended
wings, that, however, are difficult to detect in many cases because
of line blending.

In Fig.~\ref{Fig_CrII_prof} we show in more detail the two most
intense lines of Cr {\scriptsize II} multiplet 44. The line profiles
are well reproduced with a multiple Gaussian fit, with two components
in emission with a FWHM of 45 and 180 $\mathrm{km\, s^{-1}}$ FWHM
respectively, having the same central wavelength, and one absorption
with a FWHM of 25-30 $\mathrm{km\, s^{-1}}$, slightly displaced to
the red. (See the infalling material model discussed by Israelian
and Musaev.)

Strong absorption components have also been observed in the Na {\scriptsize I}
D doublet since a long time, and were discussed by de Winter et al.
(\cite{dWvdA97}). In our spectra the Na {\scriptsize I} D doublet
shown in Fig.~17 underwent dramatic variation in both intensity and
position of the emission components, as well as in the intensity of
the central absorption (its position is stable within the fitting
precision). This behaviour is not unexpected because Israelian et
al. (\cite{Ia96}) report variability for these lines within 2-3 days.
Due to the intense absorption, the narrow emission component, if present,
cannot be found by the fitting procedure. The velocity and fluxes
of absorption and broad emission components are given for the three
epochs in Table.~\ref{Tab_NaI}.

\begin{table}[h]

\caption{Flux (in $10^{\textrm{-}13}\,\mathrm{erg\, cm^{2}\, s^{-1}}$) and
velocity (in $\mathrm{km\, s^{-1}}$) variations of the Na I D1 and
D2 components.\label{Tab_NaI}}

\begin{tabular}{p{1.5cm}p{0.75cm}p{0.6cm}p{0.75cm}p{0.6cm}p{0.75cm}p{0.6cm}}
\hline 
&
1995 \foreignlanguage{frenchb}{}flux&
1995 vel&
1999 flux&
1999 vel&
2002 flux&
2002 vel\tabularnewline
\hline
Na I D1 a&
-0.88&
31&
-1.85&
24&
-2.60&
24\tabularnewline
Na I D1 e&
2.77&
48&
4.84&
4&
2.86&
9\tabularnewline
Na I D2 a&
-1.26&
31&
-2.05&
22&
-2.55&
22\tabularnewline
Na I D2 e&
3.34&
48&
5.24&
5&
2.88&
6\tabularnewline
\end{tabular}
\end{table}

As for the Fe {\scriptsize II} lines, a central absorption is clearly
present in the line profiles at $\lambda$~4508.288 $\textrm{Å}$
in Fig.~\ref{Parts_spec} and Fig.~\ref{Prof_blend_4508} and at
$\lambda$~4520.224 $\textrm{Å}$, 4549.474 $\textrm{Å}$ and 4583.837
$\textrm{Å}$ in Fig.~\ref{Parts_spec}. The absorptions, like the
Cr {\scriptsize II} ones, are red shifted an average of 3 $\mathrm{km\, s^{-1}}$
with respect to the central wavelength of the emission; the shift
is the same for all the observing runs (Table.~\ref{Tab_Mean-radial-velocities}).

The same set of components than for Cr {\scriptsize II} lines with
similar parameters is present in the Fe {\scriptsize II} lines (FWHM
of 45 and 180 \textbf{$\mathrm{km\, s^{-1}}$}), while the {[}Fe {\scriptsize II}{]}
display narrow and broad emissions with smaller FWHM of 25 and 100
\textbf{$\mathrm{km\, s^{-1}}$} , respectively (see Fig.~\ref{Prof_blend_4508}~for
highly blended lines and Fig.~\ref{Prof_blend_5273_5276}~for more
separated ones).

As the FWHM of the Fe {\scriptsize II} absorption components FWHM
are similar to the Cr {\scriptsize II} ones ($\simeq$30 $\mathrm{km\, s^{-1}}$),
these components probably originate from the same region.

We postulate that in HD45677 the double-peaked Fe{\scriptsize ~II}
line structure reported in the literature is in fact due to the same
type of absorption and originates from material in the star's line
of sight. Their redward displacement can be interpreted as due to
the infall of the absorbing material. This result is in agreement
with the observations and the conclusions of Grady et al. (\cite{GBS93})
and Israelian and Musaev (\cite{IM97}). 

The two emission components clearly originate from regions of different
dynamical behaviour, a possible model of the environment of FS CMa
will be given in Sec. \ref{ssec:Model}.

Based on the above results, we have fitted all, strong and weak, metallic
permitted lines identified in our spectra with three Gaussian line
components, central absorption narrow emission and broad emission.
The forbidden lines were fitted with two emission components. The
same central wavelength was assumed for the two emission components.
The derived line fluxes, FWHMs and central radial velocities are given
in Table.12 (online material). For the line fluxes we estimated
from the fitting procedure an error from 15\% for the strongest lines
to 50\% for the weakest lines.

\begin{table}[H]

\caption{Mean radial velocities (in $\mathrm{km\, s^{-1}}$) of the most numerous
species, absorption and emission of Cr {\scriptsize II,} Fe {\scriptsize II},
and {[}Fe {\scriptsize II}{]} emission. The number of lines is given
in parentheses.\label{Tab_Mean-radial-velocities}}

\begin{tabular}{p{1.0cm}p{2.0cm}p{2cm}p{2.0cm}}
\hline 
&
1995&
1999&
2001\tabularnewline
\hline
Cr {\scriptsize II} a&
$25.4\pm1.3$(10)&
$25.4\pm1.4$(11)&
$21.8\pm$0.8(4)\tabularnewline
Cr {\scriptsize II} e&
$23.6\pm1.3$(11)&
$23.1\pm1.2$(11)&
$18.6\pm2.0$(4)\tabularnewline
Fe {\scriptsize II} a&
$29.4\pm0.7$(58)&
$28.2\pm0.6$(59)&
$24.2\pm0.6$(35)\tabularnewline
Fe {\scriptsize II} e&
$25.8\pm0.5$(69)&
$24.5\pm0.4$(64)&
$21.6\pm0.5$(47)\tabularnewline
{[}Fe {\scriptsize II}{]}&
$21.5\pm0.5$(85)&
$22.5\pm0.4$(77)&
$21.4\pm0.3$(32)\tabularnewline
\end{tabular}
\end{table}

The mean heliocentric radial velocity of the narrow emission and central
absorption components of Cr{\scriptsize ~II}, Fe{\scriptsize ~II}
and {[}Fe{\scriptsize ~II}{]} in the three epochs are given in Table~\ref{Tab_Mean-radial-velocities}.

We compared the fluxes of the narrow emission components of Fe~\textsc{\scriptsize II}
and {[}Fe{\scriptsize ~II}{]} lines in 1995 and in 2001-2002 with
those of 1999. 

Between 1995 and 2001-2002 the fluxes of the permitted lines are the
same within the measurements errors, except for the weaker lines which
are fainter in 1995. In December 2001-January 2002 the forbidden lines
have weakened by somewhat less than 30\% with respect to 1995 and
1999. As for the broad components, their larger measurement errors
only allow to put an upper limit of 30\% to their variation during
our observing period.

In the following we shall analyse the broad and narrow emission components
of these species, with special regard to the February 1999 data which
cover a wider spectral range.

\section{Analysis of the emission line fluxes\label{Emission lines}}

The spectrum of HD 45677 is rich in emission lines (see online material 
table 13), whose strength can be used to derive information about the 
physical parameters and geometry of the circumstellar region. Here we analyse
the intensity of the permitted and forbidden ionized iron emission
lines and permitted ionized chromium emission lines in a statistical
way, the Self--Absorption--Curve Method (SAC), following the procedure
described by Friedjung \& Muratorio (\cite{FM87}), and used among
others by Muratorio et al. (\cite{MVFBR92}), Viotti \& al. (\cite{VRB99}),
Viotti et al. (\cite{VSBR00}), Kotnik-Karuza et al. (\cite{KFS02})
and van den Ancker et al. (\cite{vdABTAD04}). Details on the SAC
method are given in the {\it SAC Manual} (Baratta et al. \cite{Ba98},
\textbf{}available at www.rm.iasf.cnr.it /ftp/uvspace/). 

The Self absorption Curve Method has been discussed by Kastner (\cite{K99}),
in the framework of the escape probability formalism of radiative
transfer. He found, with such a formalism, that for a static source
having negligible differential velocities, the method is valid up
to optical thicknesses of about 10. The self absorption curve we use
is very similar to that found from Kastner's approach. As we shall
see, the use of a static model can be justified for one of the emission
line components, while most of the Fe {\scriptsize II} emission lines
examined, do not have an optical thickness larger than 10.

The method makes a certain number of assumptions, as follows :

(a) The populations of different levels belonging to the same spectroscopic
term are approximately proportional to their statistical weights.

(b) The self absorption of lines of different multiplets is always
the same function of the optical thickness. In the optical region
studied by us, our analysis is mainly based on lines whose lower levels
are both even and metastable, while the upper levels are odd. The
processes of excitation are supposed to be similar. If such a situation
exists, we may expect that effects of inhomogeneities in the emitting
region, affect the different lines in almost the same way.

(c) Selective excitation mechanisms, and particularly pumping of populations
of upper levels of emision lines because of wavelength coincidences
with other strong emission lines, are fairly infrequent and can be
neglected for most lines. Such an assumption would be quite dangerous
in the ultraviolet wavength range (Eriksson et al. \cite{EJW04}).

Other assumptions can often be made, as in this paper, when applying
the method:

(d) The populations of different collisionally excited even lower
terms of $\mathrm{Fe^{+}}$ follow roughly Boltzmann's law. In fact
one can expect LTE at least for those having a lower excitation potential.
The only relevant calculations of Verner et al. (\cite{VGBJID02})
for emission by the B+D Weigelt blobs of $\eta$~Car, with an estimated
electron density of the order of only $10^{6}cm^{-3}$, suggest that
Boltzmann's law is obeyed up to 2.3 ev, with deviations as a function
of excitation potential for higher metastable levels. Our lines are
formed in a circumstellar region, with a presumably much higher electron
density and larger collisional excitations and de-excitation rates
of the metastable levels, as indeed is consistent with the determined
column densities and limits on the size of the line emitting region,
given below (see Section 5.2, for what may happen, if this assumption
is relaxed.).

(e) The populations of the upper odd levels are not in LTE, but we
suppose that they tend to cluster around a Boltzmann law like distribution.
Our analysis of HD 45677 is consistent with such an assumption for
levels, which are not highly excited (see below), irrespective of
what the dominant excitation mechanism may be. Let us note that the
Verner et al (\cite{VGBJID02}) $\eta$~Car analysis, indicates departure
coefficients, which tend to be mostly a function of the excitation
potential.

\subsection{Fe {\scriptsize II}  permitted emission lines\label{Fe-II-permitted-emission-lines}}

\begin{figure}[h]
\includegraphics[%
  bb=40bp 80bp 520bp 750bp,
  clip,
  width=65mm,
  keepaspectratio,
  angle=270]{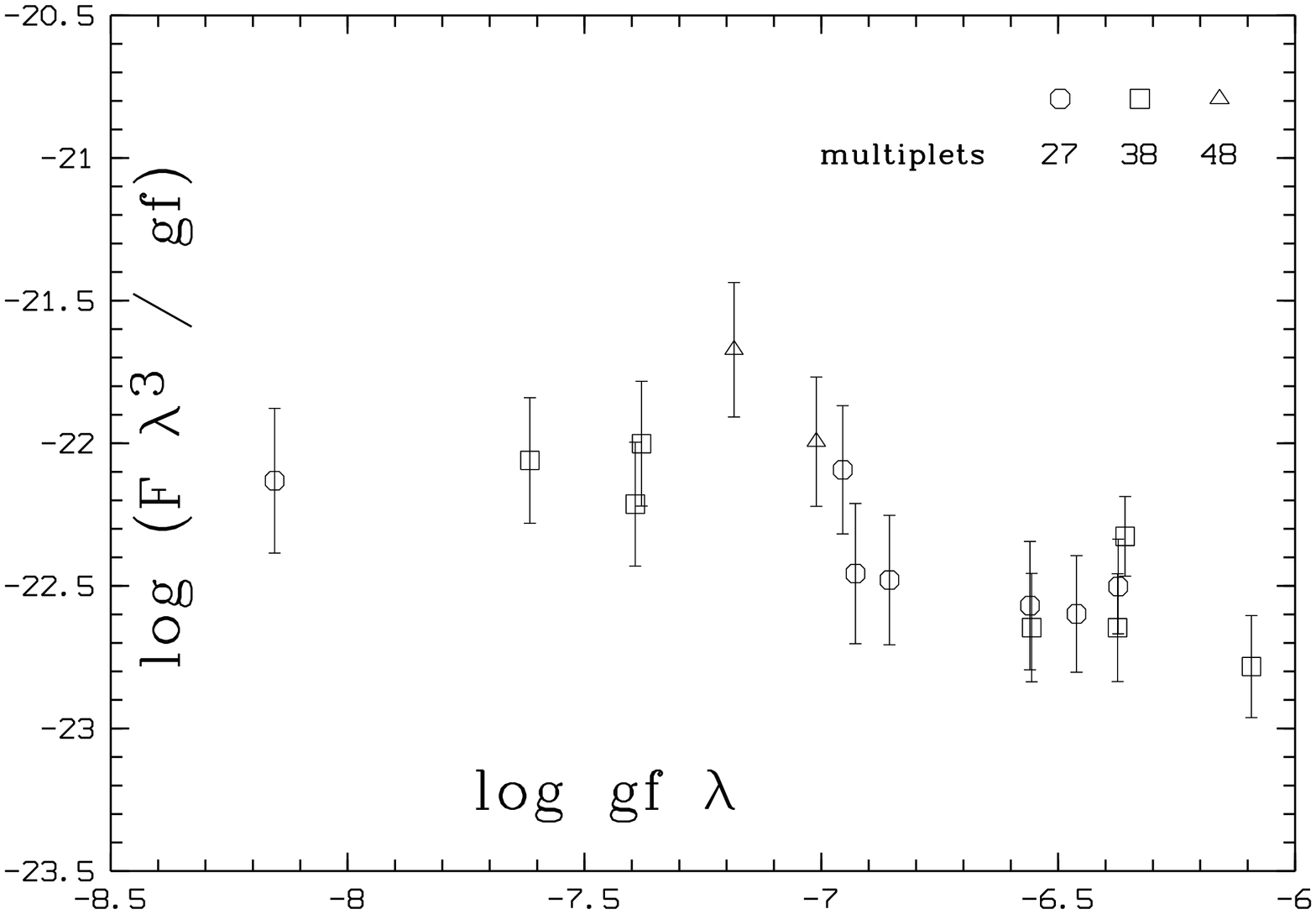}

\includegraphics[%
  bb=40bp 80bp 520bp 750bp,
  clip,
  width=65mm,
  keepaspectratio,
  angle=270]{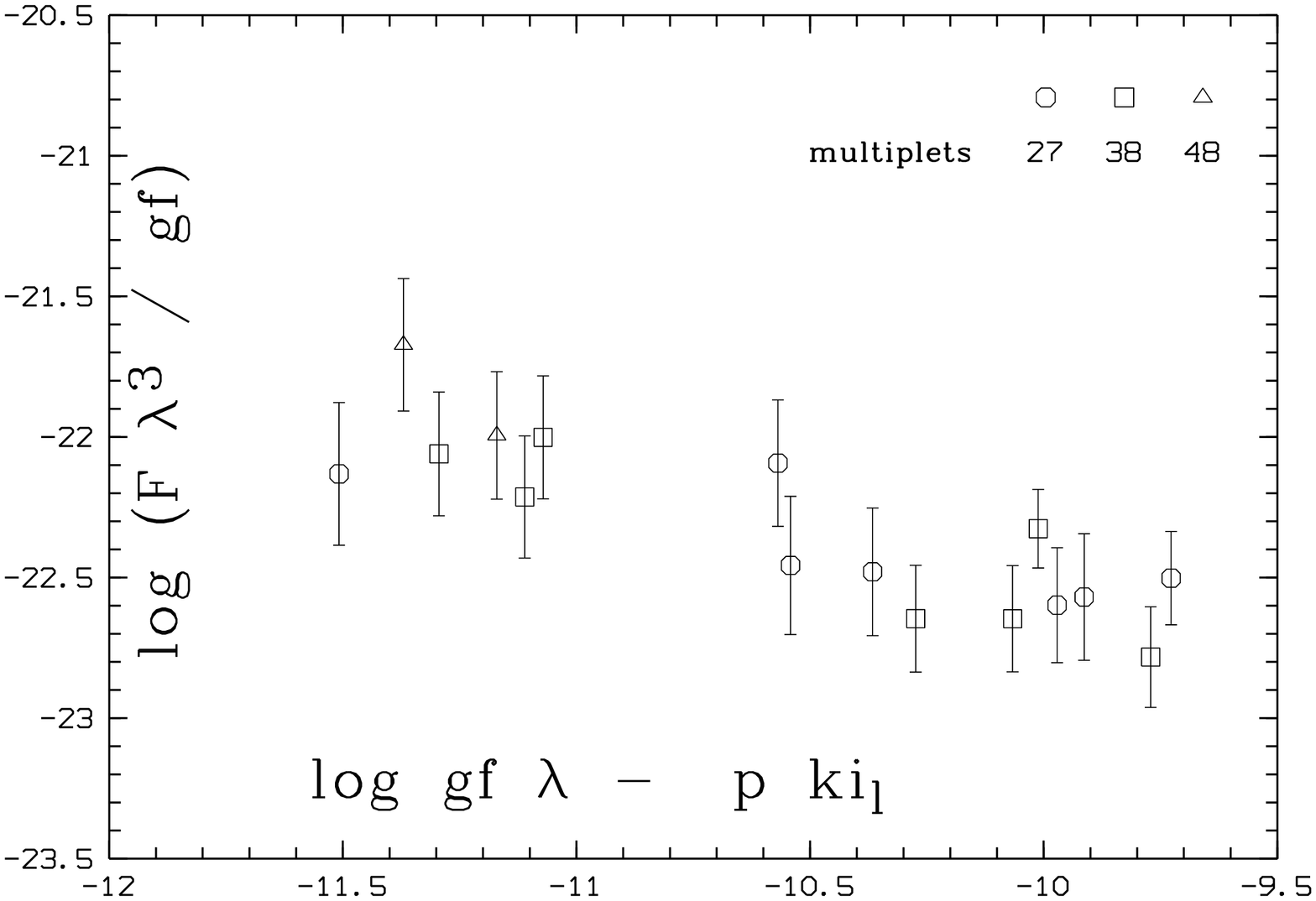}

\caption{Upper panel : Log normalized fluxes versus $\log\, gf\lambda$ plots
for the narrow components of Fe {\scriptsize II} multiplets 27 (circle), 
38 (square),48 (triangle) with common upper term $\mathrm{z^{4}D^{0}}$ in
the 1999 spectrum of HD 45677. Lower panel : log normalized fluxes
versus $\mathrm{\log\, gf\lambda-p\chi_{l}}$ plots of multiplets
27, 38 and 48. The multiplets are horizontally shifted according to
the population ratio of their lower terms by p = 1.3.\label{Individual-Fe-II_SAC}}
\end{figure}

We first consider the narrow emission components of Fe{\scriptsize ~II}.
For the analysis we adopt Kurucz' (\cite{K93}) oscillator strengths.
In order to look for self--absorption effects, we have analysed the
narrow component line fluxes of individual Fe {\scriptsize II} multiplets
in the {[}$\mathrm{\log\, gf\lambda,\log\,\frac{F\lambda^{3}}{gf}}${]}
diagram, where the quantity $\mathrm{\frac{F\lambda^{3}}{gf}}$ is
the \textit{normalized flux} of an individual emission line, and $\mathrm{gf\lambda}$
is proportional, for a given multiplet, to the optical thickness at
the line centre (see Viotti et al. (\cite{VSBR00})). This procedure
requires of course the first of the above assumptions (a).

In the upper panel of Fig. \ref{Individual-Fe-II_SAC} we show the
diagram for multiplets 27, 38 and 48 that have $z^{4}D^{0}$ as common
upper term. In the diagram, if the lines were optically thin, the
normalized emission line flux should be constant within a multiplet.
We note on the contrary that the strongest lines of multiplets 27
and 38 have a normalized flux lower than that of the weaker multiplet
lines. This indicates that they are affected by a certain amount of
self--absorption. It should also be noticed that we have not found
for any multiplet any systematic trend, within the measurement errors
of the line fluxes, with respect to the quantum number J, which would
suggest a contradiction with assumption (a), about level populations
within the upper and lower terms of each multiplet\textbf{.} Anyhow,
a thourough study of this effect will be difficult for such complex
line profiles, even with much higher quality material. 

The bend of these diagrams helps to determine the relative population
of the lower terms of the multiplets, by their horizontal shifts with
respect to each other by an amount equal to the logarithm of this
population ratio (e.g. Kotnik-Karuza et al. \cite{KFS02}). This has
been performed in the {[}$\mathrm{\log\, gf\lambda-p\chi_{l},\log\,\frac{F\lambda^{3}}{gf}}${]}
diagram (lower part of Fig. \ref{Individual-Fe-II_SAC}) where the
multiplets have been overlapped assuming a power law with exponent
-p. A value of $\mathrm{p=}1.3\pm0.4\, eV^{-1}$) is derived for the
exponent of the power law of the lower terms population of the multiplets
27, 38 and 48. 

Similarly, by comparing the 1999 diagrams for multiplets 28, 37 and
49, having $z^{4}F^{0}$ as common upper term, we derive a mean lower
level population gradient of $\mathrm{p=1.0\pm0.4\, eV^{-1}}$.

In general these results are fairly uncertain because of the small
energy range of the lower terms, and the same is the case of multiplets
with common lower term, which cannot be properly used to derive the
mean excitation of the upper terms because of the small energy range
of our data. In order to overcome this problem, we have assumed that
all the multiplets fit a common theoretical self--absorption curve
$\mathrm{Q\left(\tau\right)}$, and have used the least squares method
to determine the parameters p, $\mathrm{X_{c}}$, q, $\mathrm{Y_{c}}$,
needed to overlap the lines of all multiplets to the theoretical curve
in an {[}$\mathrm{X-X_{c},Y-Y_{c}}${]} diagram, where they correspond
to the following physical definitions :

\begin{eqnarray}
\mathrm{X} & = & \log\,\mathrm{gf\lambda}-\mathit{\mathsf{\mathit{\mathit{p}}(\chi_{l}-\chi_{c})}}\label{X}\end{eqnarray}

\begin{equation}
\mathrm{Y=\log\frac{F\lambda^{3}}{gf}+\mathit{q}(\chi_{u}-\chi_{c})}\label{Y}\end{equation}

\begin{equation}
\mathrm{X_{c}=1.576-\log\,\frac{N_{c}}{g_{c}}+\log\, v_{c}}\label{X0}\end{equation}

\begin{equation}
\mathrm{Y_{c}=\log\frac{S'}{d^{2}}+\log\frac{N_{c}}{g_{c}}+w-\log\frac{2\pi e^{2}h}{m_{e}}}\label{Y0}\end{equation}

and~ $\log\frac{2\pi e^{2}h}{m_{e}}=16.977$ ~in cgs units.

The quantities $\chi_{l}$ and $\chi_{u}$ are the excitation potentials
(in eV) of the lower and upper levels of each transition, $\chi_{c}$
is a characteristic excitation potential taken different from 0 eV
in the case of Fe~\textsc{\scriptsize II} permitted lines (and equal
to 2.75 eV) in order to avoid problems \textcolor{black}{of the} populations
of the higher metastable levels, whose possible large departures from
LTE are pointed out by Verner et al. (\cite{VGBJID02}). The unknown
parameters $\mathrm{p}$ and $\mathrm{q}$ describe the mean population
distributions of the lower and upper terms, respectively (Muratorio
et al. \cite{MVFBR92}). In this procedure we assume the first of
the above assumtions (a)\textcolor{black}{,} as verified in each multiplet
\textbf{{[}$\mathrm{\log\, gf\lambda,\log\,\frac{F\lambda^{3}}{gf}}${]}}
diagram, and that \textcolor{red}{}\textcolor{black}{all the lower
and upper levels} follow a Boltzmann--type law with excitation temperatures
$\mathrm{T_{l}}$ and $\mathrm{T_{u}}$ of $5040/\mathrm{p}$ and
$5040/\mathrm{q}$, respectively. \textbf{}

The zero points $\mathrm{X}_{c}$ and $\mathrm{Y}_{c}$, are related
to the $\mathrm{Fe^{+}}$ column density $\frac{\mathrm{N}_{c}}{g_{c}}$
at the arbitrary choosen level $\chi_{c}$, taken to be equal to 2.75
eV for Fe~\textsc{\scriptsize II} permitted lines, and to the projected
extension of the emitting volume $\mathrm{S}'$ ($\mathrm{S'=S}\cos i$),
through the equations \ref{X0} and \ref{Y0} (Friedjung \& Muratorio
\cite{FM87}; Baratta et al. \cite{Ba98}). The column density $\frac{\mathrm{N}_{c}}{g_{c}}$
also depends on the parameter $v_{c}$ (eq. (\ref{X0}), which is
the velocity broadening of the opacity profile (see Section~\ref{ssection:Column-density}).
For the theoretical SAC function $\mathrm{Q\left(\tau\right)}$ we
adopted the analytical expression for a Gaussian line profile and
an homogeneous emitting envelope given by Friedjung \& Muratorio (\cite{FM87}).

All the quantities are in cgs units. The term w in eq.~\ref{Y0}
describes the overall deviation of the population of the upper (odd)
levels of the Fe~{\scriptsize II} transitions with respect to the
population of the lower (even) ones. If we assume that the odd and
even level population laws link up at an intermediate $\overline{\chi}$
, the term turns out to be equal to $w=(\overline{\chi}-\chi_{c})\left(q-p\right)$,
where for the case of Fe~{\scriptsize II} one can take $\overline{\chi}=4.72\, eV$.
Since we expect the odd terms to be largely underpopulated with respect
to the even ones (see e.g. Fig. 13 in Verner et al. (\cite{VGBJID02})),
this value of w should be taken as an upper limit, which corresponds
to a lower limit to S'.

\begin{table}[h]

\caption{SAC fit parameters for the permitted Fe {\scriptsize II} narrow component
lines. p ($\mathrm{eV^{-1}}$), $\mathrm{X}_{c}$, q ($\mathrm{eV^{-1}}$)
and $\mathrm{Y}_{c}$ are fitting parameters,$\chi^{2}$ is a measure
of the fitting precision.\label{TAB-fit_N}}

\begin{tabular}{|c|c|c|c|}
\hline 
year&
1995&
1999&
2001\tabularnewline
\hline
\hline 
p&
$1.32\pm0.33$&
$1.28\pm0.23$&
$0.94\pm0.31$\tabularnewline
\hline 
$\mathrm{X}_{c}$&
$-7.27\pm0.09$&
$-7.52\pm0.08$&
$-7.37\pm0.11$\tabularnewline
\hline 
q&
$1.79\pm0.19$&
$1.60\pm0.16$&
$1.28\pm0.17$\tabularnewline
\hline 
$\mathrm{Y}_{c}$&
$-17.00\pm0.54$&
$-17.40\pm0.47$&
$-18.37\pm0.47$\tabularnewline
\hline 
N lines&
50&
50&
28\tabularnewline
\hline 
red.$\chi^{2}$&
0.65&
0.45&
0.34\tabularnewline
\hline
\end{tabular}
\end{table}

\begin{figure}[h]
\includegraphics[%
  bb=50bp 110bp 510bp 740bp,
  clip,
  width=7cm,
  height=9cm,
  angle=270]{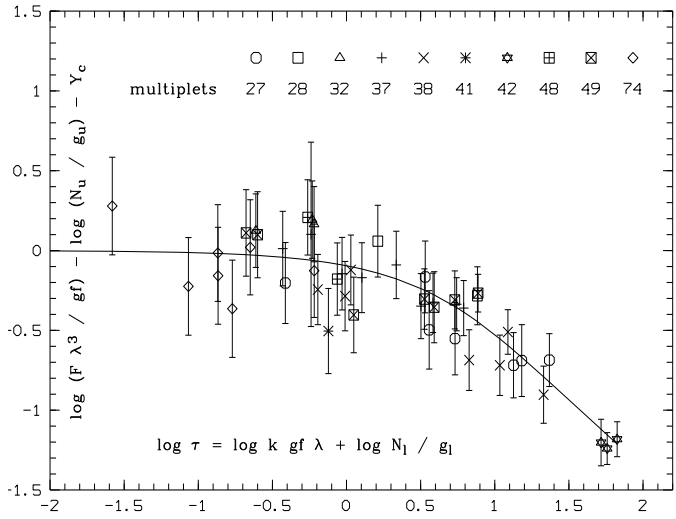}

\caption{Global SAC of the narrow emission components in the 1999 spectrum
of HD 45677. Results from the least squares overlapping of multiplets
27, 28, 32, 37, 38, 41, 42, 48, 49 and 74. The representative points
are shifted horizontally and vertically according to a least squares
fitting procedure by log level population corresponding to excitation
temperatures of 3900 K and 3150 K for the lower and upper terms respectively.\label{SACN99_1.09_1.60}}
\end{figure}

Fig.~\ref{SACN99_1.09_1.60} shows the result of the least squares
fit of the dereddened fluxes (with error bars) of 49 narrow components
of the Fe {\scriptsize II} lines measured in the 1999 spectrum of
HD 45677. The high excitation ($\chi_u > 7.5$ eV) Fe {\scriptsize II}
lines have not been included in the fit, and will be discussed separately
in Section \ref{High-excitation-FeII}.

As expected, in Fig.~\ref{SACN99_1.09_1.60} multiplet 74 emission
lines lie on the optically thin branch of the SAC, while the very
strong lines of multiplet 42 lines are the most optically thick ones
(central line opacity of about 70). Note that many multiplets lie
around the bend of the SAC ($\log\tau\,\simeq\,0$).

The fit gives a mean lower level population p corresponding to an
excitation temperature $\mathrm{T}_{l}$ of $3900_{-600}^{+900}\,\mathrm{K}$,
in fairly good agreement with the values previously found by considering
two groups of multiplets having common upper terms. The resulting
mean upper level population parameter q corresponds to a mean excitation
temperature $\mathrm{T}_{u}$ of $3150_{-300}^{+350}\,\mathrm{K}$,
much smaller than that of the lower levels. Let us note that the modelling
of Verner et al. (\cite{VGBJID02}) for two of the Weigelt blobs of
$\eta$~Car predicted electron temperatures going down to 5000~K.

We have applied this analysis also to the narrow Fe {\scriptsize II}
emission components in the 1995 and 2001--2002 spectra of HD 45677.
The derived best fit parameters for the three epochs are summarized
in Table.~\ref{TAB-fit_N}. One may note in the table that the derived
$\mathrm{Y_{c}}$ parameters are very different from one epoch to
another. This cannot be totally ascribed to a variation of the size
of the Fe {\scriptsize II} emitting region, to which the $\mathrm{Y_{c}}$
parameter is related, since, as noted in Sec.~\ref{sub:Line-components},
the spectrum did not largely vary during 1995-2002. Actually, the
fitting value of $\mathrm{Y_{c}}$ is correlated with the q parameter,
so that a small error on q results in a large change in $\mathrm{Y_{c}}$.
In view of this point, as also discussed above, in this paper we shall
mostly concern ourselves with the 1999 spectrum of HD 45677, and leave
the problem of the spectral variation to a forthcoming paper.

\begin{figure}[h]
\includegraphics[%
  bb=50bp 110bp 510bp 740bp,
  clip,
  width=7cm,
  height=9cm,
  angle=270]{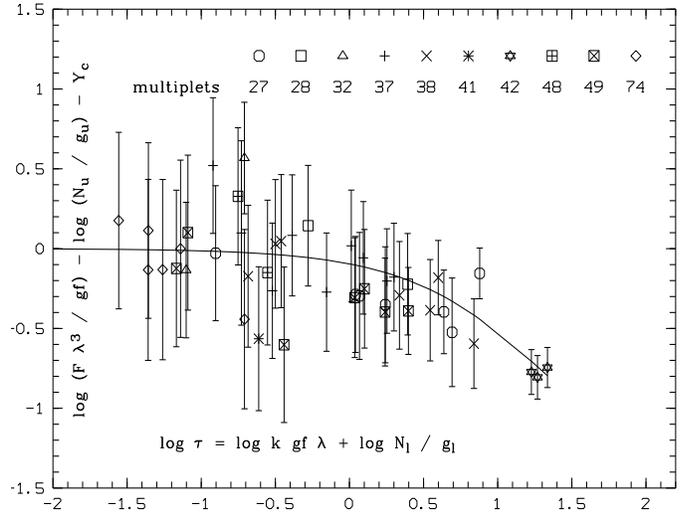}

\caption{SAC of broad components of multiplets 27, 28, 32, 37, 38, 41, 42,
48, 49 and 74 of Fe {\scriptsize II} in 1999. The representative points
are shifted horizontally and vertically by log~level~population
corresponding to excitation temperatures of respectively $3900$ K
and $3400$ K\label{SACW99_1.09_1.61}}
\end{figure}

The same procedure has been applied to the Fe {\scriptsize II} broad
emission components components. Fig.~\ref{SACW99_1.09_1.61} shows
the global SAC of the Fe {\scriptsize II} broad emissions in the 1999
spectrum of HD 45677 obtained from the least squares overlapping of
ten Fe {\scriptsize II} multiplets. The fluxes of the broad components
are less precisely determined than the narrow component ones, and
the scatter of their SAC curves is higher, respectively.

\begin{table}[h]

\caption{SAC fit parameters for the permitted Fe {\scriptsize II} broad component
lines. See Table. \ref{TAB-fit_N} and text for definitions. \textcolor{black}{{*}
parameter taken equal to the narrow components one as no convergence
is obtained.}\label{TAB-fit_W}}

\begin{tabular}{|c|c|c|c|}
\hline 
year&
1995&
1999&
2001\tabularnewline
\hline
\hline 
p&
$1.32*$&
$1.28*$&
$0.94*$\tabularnewline
\hline 
$\mathrm{X}_{c}$&
$-6.97\pm0.14$&
$-7.03\pm0.11$&
$-7.29\pm0.18$\tabularnewline
\hline 
q&
$0.99\pm0.29$&
$1.50\pm0.26$&
$1.53\pm0.26$\tabularnewline
\hline 
$\mathrm{Y}_{c}$&
$-19.35\pm0.82$&
$-17.84\pm0.75$&
$-17.84\pm0.74$\tabularnewline
\hline 
N lines&
49&
49&
28\tabularnewline
\hline 
red.$\chi^{2}$&
0.45&
0.41&
0.33\tabularnewline
\hline
\end{tabular}
\end{table}

The fitting parameters of the Fe {\scriptsize II} broad emission components
are given in Table \ref{TAB-fit_W}.

The SAC analysis of the broad components in 1995 and 2001-2002 is
more uncertain, due to the lower spectral resolution in 1995, which
results in a less precise determination of the broad component fluxes,
and to the smaller set of data in 2001--2002. \textcolor{black}{Despite
the lack of convergence for the p value in 1999}\textbf{,} the parameters,
summarized in Table.~\ref{TAB-fit_W}, appear fairly similar to the
ones derived from the narrow emission components except for the 1995
data, despite the too small q value derived.

\subsection{High excitation Fe {\scriptsize II} permitted lines\label{High-excitation-FeII}}

\begin{table}[h]

\caption{For each high excitation line we give : (1) the laboratory wavelength,
(2) upper and lower term of the transition, (3) upper and lower term
excitation potential, (4) logarithm of oscillator strength, (5) log
ratio of observed flux to calculated flux according to the lower excitation
lines parameters.\label{HEL_Tab}}

\selectlanguage{frenchb}
\begin{tabular}{|m{0.17\columnwidth}|m{0.20\columnwidth}|m{0.11\columnwidth}|c|m{0.12\columnwidth}|}
\hline 
\selectlanguage{frenchb}
Wavelength
\selectlanguage{english}&
\selectlanguage{frenchb}
upper Term

lower Term
\selectlanguage{english}&
\selectlanguage{frenchb}
$\chi_{up}$

$\chi_{low}$
\selectlanguage{english}&
\selectlanguage{frenchb}
$\log\, gf$
\selectlanguage{english}&
\selectlanguage{frenchb}
$\log\,\frac{F_{obs}}{F_{calc}}$
\selectlanguage{english}\tabularnewline
\hline
\hline 
\selectlanguage{frenchb}
6158.019
\selectlanguage{english}&
\selectlanguage{frenchb}
(3G)4p x4F

(3F)4s c4F
\selectlanguage{english}&
\selectlanguage{frenchb}
8.23

6.22
\selectlanguage{english}&
\selectlanguage{frenchb}
-3.09
\selectlanguage{english}&
\selectlanguage{frenchb}
4.1
\selectlanguage{english}\tabularnewline
\hline 
\selectlanguage{frenchb}
6160.755
\selectlanguage{english}&
\selectlanguage{frenchb}
(3H)4p z4H

(1F)4s c2F 
\selectlanguage{english}&
\selectlanguage{frenchb}
7.58

5.57
\selectlanguage{english}&
\selectlanguage{frenchb}
-3.63
\selectlanguage{english}&
\selectlanguage{frenchb}
3.5
\selectlanguage{english}\tabularnewline
\hline 
\selectlanguage{frenchb}
6233.534
\selectlanguage{english}&
\selectlanguage{frenchb}
d5 4s2 c4D

(5D)4p z4F
\selectlanguage{english}&
\selectlanguage{frenchb}
7.47

5.48
\selectlanguage{english}&
\selectlanguage{frenchb}
-2.94
\selectlanguage{english}&
\selectlanguage{frenchb}
2.7
\selectlanguage{english}\tabularnewline
\hline 
\selectlanguage{frenchb}
6248.898
\selectlanguage{english}&
\selectlanguage{frenchb}
d5 4s2 c4D

(5D)4p z4D
\selectlanguage{english}&
\selectlanguage{frenchb}
7.49

5.51
\selectlanguage{english}&
\selectlanguage{frenchb}
-2.70
\selectlanguage{english}&
\selectlanguage{frenchb}
2.9
\selectlanguage{english}\tabularnewline
\hline 
\selectlanguage{frenchb}
6317.983
\selectlanguage{english}&
\selectlanguage{frenchb}
d5 4s2 c4D

(5D)4p z4D
\selectlanguage{english}&
\selectlanguage{frenchb}
7.47

5.51
\selectlanguage{english}&
\selectlanguage{frenchb}
-1.99
\selectlanguage{english}&
\selectlanguage{frenchb}
2.6
\selectlanguage{english}\tabularnewline
\hline 
\selectlanguage{frenchb}
6371.125
\selectlanguage{english}&
\selectlanguage{frenchb}
d5 4s2 c4D

(5D)4p z4F
\selectlanguage{english}&
\selectlanguage{frenchb}
7.49

5.55
\selectlanguage{english}&
\selectlanguage{frenchb}
-3.56
\selectlanguage{english}&
\selectlanguage{frenchb}
2.8
\selectlanguage{english}\tabularnewline
\hline 
\selectlanguage{frenchb}
6383.722
\selectlanguage{english}&
\selectlanguage{frenchb}
d5 4s2 c4D

(5D)4p z4D
\selectlanguage{english}&
\selectlanguage{frenchb}
7.49

5.55
\selectlanguage{english}&
\selectlanguage{frenchb}
-2.27
\selectlanguage{english}&
\selectlanguage{frenchb}
2.7
\selectlanguage{english}\tabularnewline
\hline 
\selectlanguage{frenchb}
6385.451
\selectlanguage{english}&
\selectlanguage{frenchb}
d5 4s2 c4D

(5D)4p z4D
\selectlanguage{english}&
\selectlanguage{frenchb}
7.49

5.55
\selectlanguage{english}&
\selectlanguage{frenchb}
-2.62
\selectlanguage{english}&
\selectlanguage{frenchb}
2.9
\selectlanguage{english}\tabularnewline
\hline 
\selectlanguage{frenchb}
6442.955
\selectlanguage{english}&
\selectlanguage{frenchb}
d5 4s2 c4D

(5D)4p z4F
\selectlanguage{english}&
\selectlanguage{frenchb}
7.47

5.55
\selectlanguage{english}&
\selectlanguage{frenchb}
-2.88
\selectlanguage{english}&
\selectlanguage{frenchb}
2.9
\selectlanguage{english}\tabularnewline
\hline 
\selectlanguage{frenchb}
6455.837
\selectlanguage{english}&
\selectlanguage{frenchb}
d5 4s2 c4D

(5D)4p z4D
\selectlanguage{english}&
\selectlanguage{frenchb}
7.47

5.55
\selectlanguage{english}&
\selectlanguage{frenchb}
-2.95
\selectlanguage{english}&
\selectlanguage{frenchb}
3.0
\selectlanguage{english}\tabularnewline
\hline 
\selectlanguage{frenchb}
6482.204
\selectlanguage{english}&
\selectlanguage{frenchb}
(3G)4p x4G

(3F)4s c4F
\selectlanguage{english}&
\selectlanguage{frenchb}
8.13

6.22
\selectlanguage{english}&
\selectlanguage{frenchb}
-2.27
\selectlanguage{english}&
\selectlanguage{frenchb}
3.3
\selectlanguage{english}\tabularnewline
\hline 
\selectlanguage{frenchb}
6491.246
\selectlanguage{english}&
\selectlanguage{frenchb}
d5 4s2 c4D

(5D)4p z4D
\selectlanguage{english}&
\selectlanguage{frenchb}
7.49

5.58
\selectlanguage{english}&
\selectlanguage{frenchb}
-2.79
\selectlanguage{english}&
\selectlanguage{frenchb}
3.0
\selectlanguage{english}\tabularnewline
\hline 
\selectlanguage{frenchb}
6493.035
\selectlanguage{english}&
\selectlanguage{frenchb}
d5 4s2 c4D

(5D)4p z4D
\selectlanguage{english}&
\selectlanguage{frenchb}
7.49

5.58
\selectlanguage{english}&
\selectlanguage{frenchb}
-2.58
\selectlanguage{english}&
\selectlanguage{frenchb}
2.9
\selectlanguage{english}\tabularnewline
\hline 
\selectlanguage{frenchb}
6506.333
\selectlanguage{english}&
\selectlanguage{frenchb}
d5 4s2 c4D

(5D)4p z4F
\selectlanguage{english}&
\selectlanguage{frenchb}
7.49

5.59
\selectlanguage{english}&
\selectlanguage{frenchb}
-3.11
\selectlanguage{english}&
\selectlanguage{frenchb}
3.1
\selectlanguage{english}\tabularnewline
\hline 
\selectlanguage{frenchb}
6517.018
\selectlanguage{english}&
\selectlanguage{frenchb}
d5 4s2 c4D

(5D)4p z4D
\selectlanguage{english}&
\selectlanguage{frenchb}
7.49

5.58
\selectlanguage{english}&
\selectlanguage{frenchb}
-2.76
\selectlanguage{english}&
\selectlanguage{frenchb}
3.1
\selectlanguage{english}\tabularnewline
\hline
\end{tabular}
\selectlanguage{english}
\end{table}

\begin{figure}[h]
\includegraphics[%
  bb=50bp 110bp 510bp 740bp,
  clip,
  width=6cm,
  height=9cm,
  angle=270]{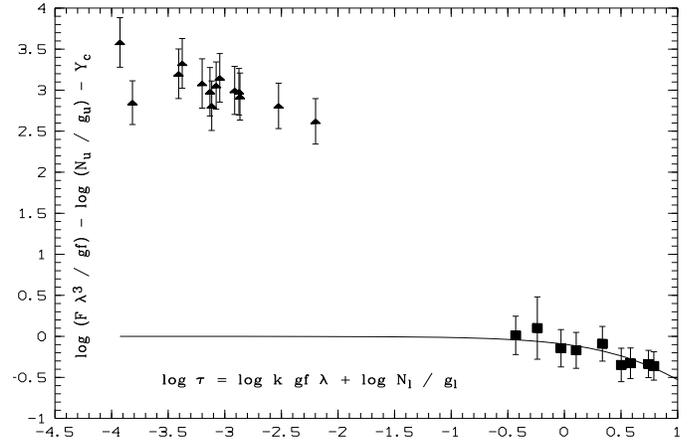}

\caption{Position of the high excitation Fe {\scriptsize II} lines (filled
triangles) in the 1999 spectrum of HD 45677 with respect to the global
SAC of Fig.~\ref{SACN99_1.09_1.60}. The multiplet 37 lines (filled
squares) are shown for comparison . \label{High_excit_SAC}}
\end{figure}

In the spectrum of HD 45677 we have identified many emission lines
arising from high energy levels ($\chi_{u}\,\geq\,7.5\:\mathrm{eV}$),
 that have not been included in the previous SAC analysis.
 Fig.~\ref{High_excit_SAC} shows the
position of the narrow components of these lines with respect to the
SAC derived from the narrow components of the lower excitation Fe
{\scriptsize II} lines. Indeed, plotting these lines in the same graph
may present a difficulty, since we cannot adopt for these lines the
same excitation temperature of the lower excitation Fe {\scriptsize II}
lines, or even define an excitation temperature for the upper levels,
as they certainly are far from equilibrium. However, it is clear from
the figure that, whatever level excitation law is adopted, the high
excitation lines are 2--3 orders of magnitude stronger than if they
were excited by the same mechanism as the lower excitation permitted
lines. The values displayed in Table~\ref{HEL_Tab} are a clear indication
of an overpopulation of the upper levels, as also appears in Fig.~\ref{level_POP}.
This will be discussed in Section \ref{sub:Level-Population}.

\subsection{Fe {\scriptsize II} \textup{forbidden emission lines \label{sub:Fe-II-forbidden}}}

\begin{figure}[h]
\includegraphics[%
  bb=50bp 110bp 510bp 740bp,
  clip,
  width=7cm,
  height=9cm,
  angle=270]{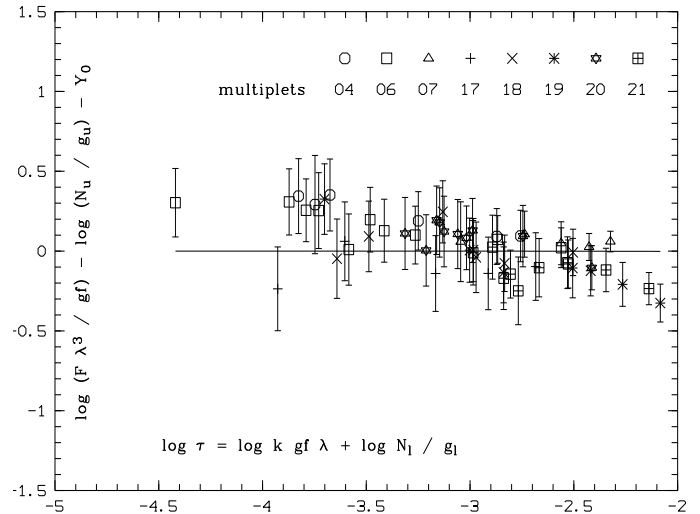}

\caption{SAC diagram of narrow components of multiplets 4F, 6F, 7F, 17F, 18F,
19F, 20F and 21F of {[}Fe {\scriptsize II}{]} in 1999. The same mean
level excitation temperature of $10\,500°$K has been used for both
lower and upper levels. The \textbf{}\textcolor{black}{observational
points are interpolated with horizontal} line.\label{SAC_O_thin-[FeII]_F}}
\end{figure}

Now we analyse the narrow and broad components of the forbidden lines
of ionized iron observed in the 1999 spectrum of HD 45677. We adopted
the {[}Fe {\scriptsize II}{]} transition probabilities of Quinet et
al. (\cite{QDZ}), and took for simplicity $\chi_{c}=0\, eV$.

\begin{table}[h]

\caption{SAC fit parameters for the {[}Fe {\scriptsize II}{]} narrow components.
q ($eV^{-1}$) and $Y_{f}$ are fitting parameters. \textbf{\textcolor{red}{}}\textcolor{black}{$\mathrm{\overline{Y}}$
is $Y_{f}$ corrected for q at a mean 3 eV excitation potential}.
$\log\overline{N}(H)$is the logarithm of the total number of Hydrogen
atom in the emitting volume.\label{TAB-fit_F}}

\begin{tabular}{|c|c|c|c|}
\hline 
year&
1995&
1999&
2001\tabularnewline
\hline
\hline 
q&
$0.44\pm0.05$&
$0.47\pm0.05$&
$0.57\pm0.08$\tabularnewline
\hline 
$Y_{\mathrm{f}}$&
$-15.51\pm0.16$&
$-15.41\pm0.13$&
$-15.20\pm0.25$\tabularnewline
\hline 
$\overline{\mathrm{Y}}$&
-16.83&
-16.82&
-16.91\tabularnewline
\hline 
$T_{f}$&
$11\,500\pm1\,000$&
$10\,500\pm1\,000$&
$9\,000\pm2\,000$\tabularnewline
\hline 
$\log\overline{N}(H)$&
$49.9\pm.2$&
$50.0\pm.1$&
$50.2\pm.3$\tabularnewline
\hline 
N lines&
85&
80&
31\tabularnewline
\hline 
red.$\chi^{2}$&
0.68&
0.82&
0.94\tabularnewline
\hline
\end{tabular}
\end{table}

The plot of {[}Fe {\scriptsize II}{]} individual multiplets in the
{[}$\mathrm{\log\, gf\lambda,\log\,\frac{F\lambda^{3}}{gf}}${]} diagram
shows a systematical downward shift of multiplets arising from higher
excitation terms, that can be used to determine their upper level
population gradient q. In the strongest multiplets there is also an
indication of the most intense lines having a smaller normalised flux.

We have overlapped the {[}Fe {\scriptsize II}{]} multiplets in the
{[}$\mathrm{X-X_{c},Y-Y_{c}}${]} plane, in the same way as we did
for permitted Fe {\scriptsize II} lines , with the least squares method,
assuming that the \textcolor{black}{ground} and lower $\mathrm{Fe^{+}}$
levels have the same excitation temperature (i.e., $p=q=5040/T_{f}$).
The derived SAC diagram shown in Fig.~\ref{SAC_O_thin-[FeII]_F}
corresponds to a best fit excitation temperature of the metastable
levels (0-4 eV) of $T_{f}=10\,500\pm1\,000\,\mathrm{K}$. Actually,
the upper {[}Fe{\scriptsize ~II}{]} levels are the same as the lower
levels of the permitted transitions treated above. Nevertheless, the
fact that two such different excitation temperatures have been derived,
added to the smaller FWHM value of the {[}Fe{\scriptsize ~II}{]},
25 $\mathrm{km\, s^{-1}}$ against 45 $\mathrm{km\, s^{-1}}$ for
Fe {\scriptsize II} permitted lines, is strongly suggestive of two
different formation regions. The results for the three epochs are
summarised in Table \ref{TAB-fit_F}. 

Let us point in Table \ref{TAB-fit_F}~that the variation in the
different epochs of the $T_{f}$ and $Y_{f}$ parameters are much
smaller than those derived for the permitted lines, probably as the
result of the smaller uncertainties. The value, at the mean upper
level excitation potential of our sample lines~$\overline{\chi_{u}}=3.0\, eV$,
of the quantity $\mathrm{\overline{Y}=-p\times\overline{\chi_{u}}+Y_{f}}$
\textcolor{black}{is smaller by a factor .8 in 2001 corresponding
approximately to the small flux decrease} between 1999 and 2001-2002
noted in Sec. \ref{sub:Line-components}.

\begin{figure}[h]
\includegraphics[%
  bb=50bp 110bp 510bp 740bp,
  clip,
  width=7cm,
  height=9cm,
  angle=270]{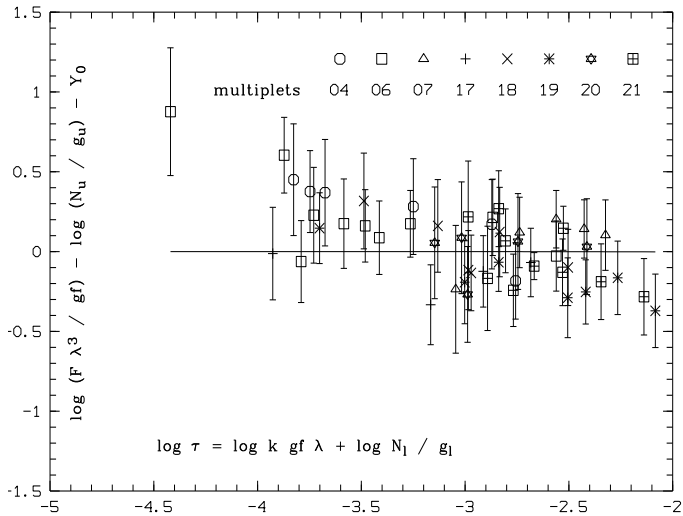}

\caption{SAC diagram of broad components of multiplets 4F, 6F, 7F, 17F, 18F,
19F, 20F and 21F of {[}Fe {\scriptsize II}{]} in 1999. The representative
points are shifted horizontally and vertically by log~level~population
corresponding to an excitation temperature of $8\,500$ K. \label{SAC-[FeII]_G}}
\end{figure}

\begin{table}[H]

\caption{SAC fit parameters for the {[}Fe {\scriptsize II}{]} broad components.\label{TAB-fit_G}}

\begin{tabular}{|c|c|c|c|}
\hline 
year&
1995&
1999&
2001\tabularnewline
\hline
\hline 
q&
$0.50\pm0.20$&
$0.64\pm0.09$&
$0.48\pm0.18$\tabularnewline
\hline 
$\mathrm{Y}_{\mathrm{f}}$&
$-15.95\pm0.60$&
$-15.38\pm0.26$&
$-15.94\pm0.53$\tabularnewline
\hline 
$\overline{\mathrm{Y}}$&
-17.45&
-17.30&
-17.38\tabularnewline
\hline 
$\mathrm{T_{f}}$&
$10\,000\pm3\,000$&
$8\,000\pm1\,500$&
$10\,500\pm3\,000$\tabularnewline
\hline 
$\log\mathrm{\overline{N}(H})$&
$49.5\pm.6$&
$50.0\pm.3$&
$49.5\pm.5$\tabularnewline
\hline 
N lines&
61&
63&
28\tabularnewline
\hline 
red.$\chi^{2}$&
2.0&
1.23&
1.45\tabularnewline
\hline
\end{tabular}
\end{table}

As for the broad components of the {[}Fe {\scriptsize II}{]} emission
lines, the strength of the better measured lines seems to suggest
by itself an emitting region with properties comparable to those of
the narrow components discussed above, this is confirmed by the fitting
values given in Table.~\ref{TAB-fit_G}.

\subsection{Cr {\scriptsize II} \textup{emission lines}}

\begin{figure}[h]
\includegraphics[%
  bb=50bp 110bp 510bp 740bp,
  clip,
  width=7cm,
  height=9cm,
  angle=270]{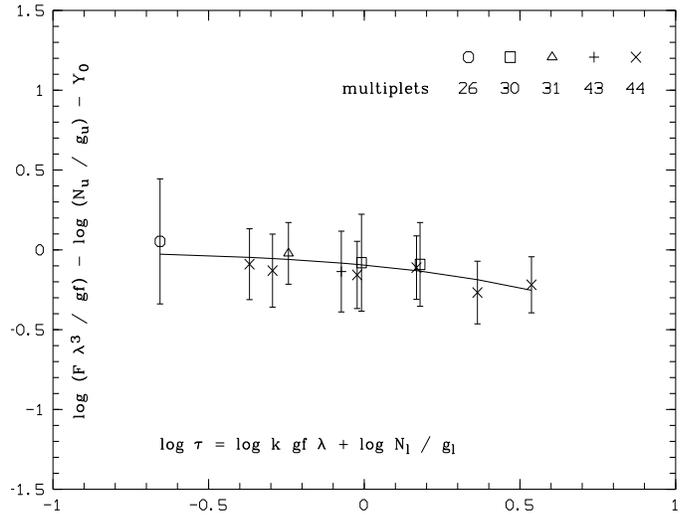}

\caption{Self Absorption Curve for the narrow components of Cr {\scriptsize II}
emission lines .\label{SAC_CrII_N}}
\end{figure}

\begin{figure}[h]
\includegraphics[%
  bb=50bp 110bp 510bp 740bp,
  clip,
  width=7cm,
  height=9cm,
  angle=270]{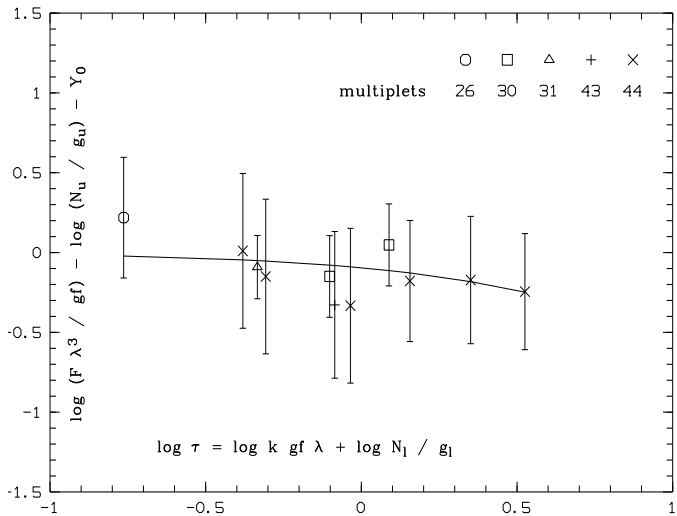}

\caption{Self Absorption Curve for the broad components of the Cr {\scriptsize II}
emission lines.\label{SAC_CrII_W}}
\end{figure}

\begin{table}[h]

\caption{SAC fit parameters for the narrow and broad component lines of $\mathrm{CrII}$
in 1999. \label{TAB-fit_Cr}}

\begin{tabular}{|c|c|c|}
\hline 
component&
narrow&
broad\tabularnewline
\hline
\hline 
p=q&
$0.82\pm0.15$&
$0.43\pm0.28$\tabularnewline
\hline 
$\mathrm{X}_{c}$&
$-8.63\pm0.60$&
$-7.03\pm1.00$\tabularnewline
\hline 
$\mathrm{Y}_{c}$&
$-18.42\pm0.98$&
$-20.87\pm1.85$\tabularnewline
\hline 
N lines&
11&
11\tabularnewline
\hline 
red.$\chi^{2}$&
0.07&
0.19\tabularnewline
\hline
\end{tabular}
\end{table}

The analysis of the narrow and broad Cr {\scriptsize II} multiplet
44 components, indicates a moderate optical thickness. Here again
we adopted $\chi_{c}=0\, eV$ in the equations defined in Section
\ref{Fe-II-permitted-emission-lines}. Due to the low number of Cr
{\scriptsize II} lines observed in the other multiplets we have derived
the SAC parameters taking the same excitation temperatures for the
lower and upper terms of both the narrow and broad components. 

The SACs plotted in Figs.~\ref{SAC_CrII_N}~and \ref{SAC_CrII_W}
correspond to the parameters given in Table.~\ref{TAB-fit_Cr}.

\section{Line profiles\label{ssec:Model}}

The two--component profile of the emission lines is suggestive of
line formation in two regions with different velocity broadening in
the observer's line--of--sight. In B{[}e{]} stars, of which HD 45677
is one important representative, low excitation emission lines, like
among others Fe {\scriptsize II} and Cr {\scriptsize II}, are generally
thought to be formed in an equatorial disk around the B star, while
the higher temperature emission lines are thought to be emitted in
a polar wind. Zickgraf (\cite{Z2003}) recently studied the kinematical
structure of B{[}e{]} star envelopes based on high resolution profiles
of permitted and forbidden emission lines, and compared them with
theoretical profiles emerging from an optically thin latitude--dependent
expanding stellar wind and an equatorial opaque dust ring. He observed
in his 1986--88 CAT--CES spectra of HD 45677 a marginal line split
of 6 $\mathrm{km\, s^{-1}}$ for the {[}O {\scriptsize I}{]} 6300~$\textrm{Å}$
emission line, while {[}Fe {\scriptsize II}{]} and {[}N {\scriptsize II}{]}
were single peaked. He interpreted his observations with line formation
in a wide opening angle polar wind seen nearly pole--on.

In HD 45677 the profile of the emission lines do not show the double
peaked structure typical of line emission from an equatorial rotating
disk seen at high inclinations. The narrow components of strongest
permitted emission lines are split by a slightly longward displaced
($\sim+3\,\mathrm{km\, s^{-1}}$) dip which we attribute instead to
absorption from matter flowing towards the star. A central absorption
with the same radial velocity is also present in the Na {\scriptsize I}
yellow doublet, and in the Mg {\scriptsize II} 4481~$\textrm{Å}$
line. No line splitting is seen in the forbidden lines. Another important
feature of HD 45677 is the narrower profile of the forbidden {[}Fe
{\scriptsize II}{]} lines, suggesting line formation in regions with
a smaller radial velocity range, like for instance the outer parts
of a rotating disk in which the inner, denser and faster rotating
parts produce the permitted emission lines. 

In HD 45677 the velocity broadening of the Si {\scriptsize II} photospheric
lines ($\leq\sim70\,\mathrm{km\, s^{-1}}$, Israelian et al. \cite{Ia96})
should indicate either a relatively low stellar rotational velocity,
or a low inclination angle $i$ of a fast rotating star. If this is
the case, the narrow emission components could arise from a dense,
fast rotating disk seen nearly face-on. If one assumes that matter
in the disk is rotating at the Keplerian velocity, the narrower {[}Fe
{\scriptsize II}{]} emission lines should be formed in the outer (and
less dense) rings of the disk. In such a framework, the broad components
of the Fe {\scriptsize II} and {[}Fe {\scriptsize II}{]} emission
lines, that, as discussed above, approximately emit a total amount
of radiative power comparable to that of the narrow components, should
be formed in region(s) with a large radial velocity range in the observer's
line-of-sight. This could be \textcolor{black}{identified} as being
a rather massive \textcolor{black}{stream} of matter flowing from
\textcolor{black}{(or to)} the stellar polar regions.

We geometrically model the emitting region of HD~45677 in order to
synthetise simultaneously both the observed profiles of the {[}Fe
{\scriptsize II}{]} emission lines and the Fe {\scriptsize II} ones. 

As discussed above, the model of a disk seen nearly edge-on proposed
for this star for instance by Swings \cite{S73} or de Winter et al.
\cite{dWvdA97} besides the fact that it would imply a too small star
rotational velocity, would generate emission line profiles with the
typical double-peaked profile totally different from the ones observed. 

\begin{figure}[h]
\includegraphics[%
  bb=40bp 120bp 550bp 790bp,
  clip,
  width=7cm,
  height=9cm,
  angle=270]{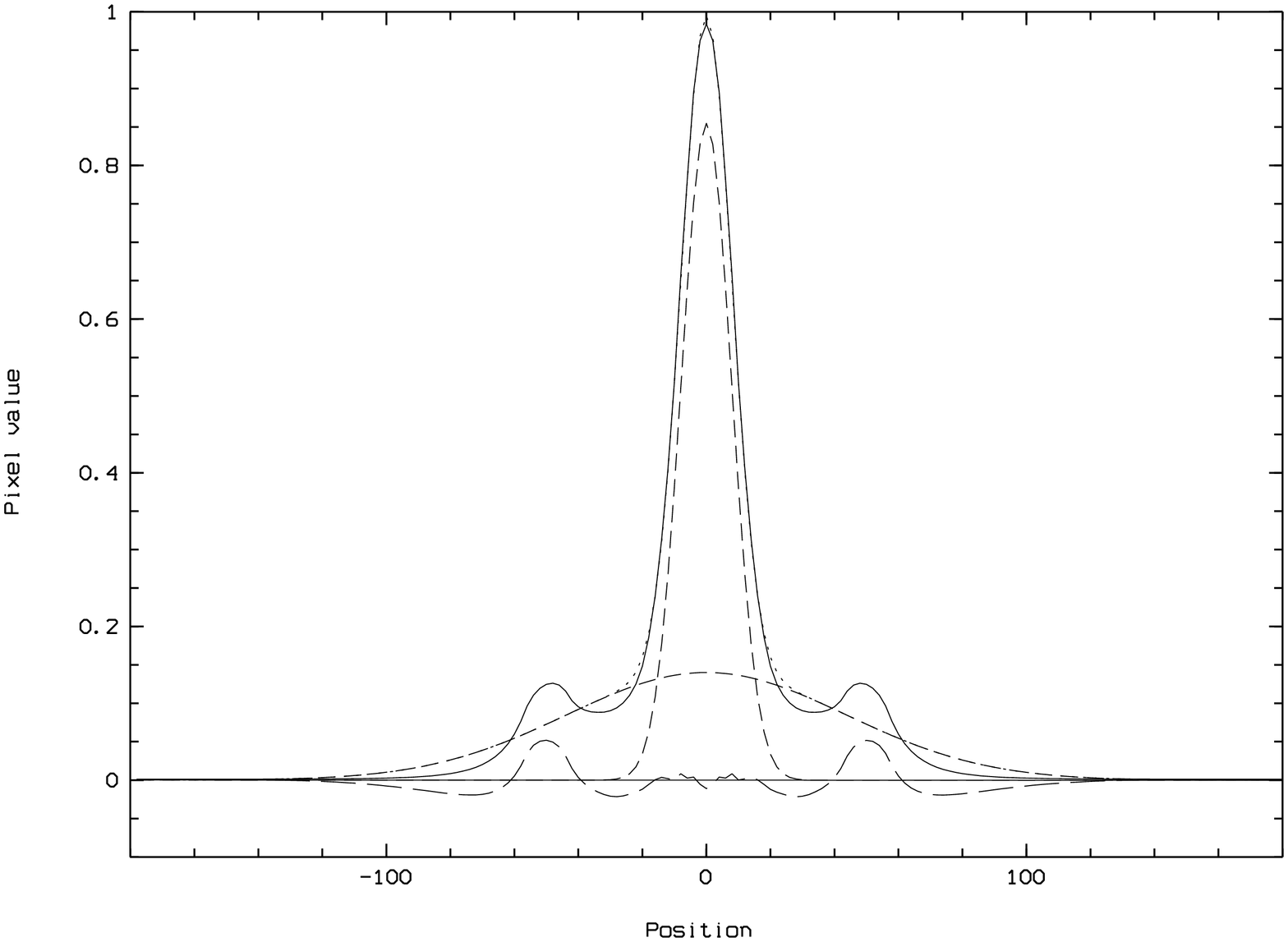}

\caption{Histogram of the emitting material radial velocities of an optically
thin disk plus polar wind. The disk is inclined 85° to the sky plane
with a total opening angle of 24°. The velocity in the disk is 150
$\mathrm{km\, s^{-1}}$ at the photospheric radius r=1, and the law
is Keplerian. The polar wind is emitted in a 30° cone with a velocity
of 100 $\mathrm{km\, s^{-1}}$ at the pole and decreases exponentialy.
In this figure and in figures \ref{Permitted_line_profile} and \ref{Forbidden_line_profile},
the synthetised profile and the continuum are plotted as a solid line,
each Gaussian component as a short dashed line, the total fit profile
as a dotted line and the residual flux as a long dashed line.}

\label{disk+wind}
\end{figure}

The bimodal model currently adopted for B{[}e{]} stars supposes different
physical conditions in the dense disk medium, where the emission lines
of the lower excitation species such as Fe {\scriptsize II} are supposed
to be produced, and in the polar wind or streams, where lines from
higher excitation species are emitted (Swings \cite{S73}, Zickgraf
\cite{Zi98}). We nevertheless took this configuration, namely an
edge-on disk and a polar wind confined in a cone having the same physical
conditions (temperature and density), into consideration. The most
similar profile was obtained with a $24^{°}$ opening angle disk inclined
$85^{°}$on the sky plane, in which the wide component is emitted,
and a $30^{°}$cone wind for wich we had to suppose a 100 $km\, s^{-1}$decreasing
velocity wind to get the narrow component. The computed profile is
shown in Fig. \ref{disk+wind}. But the result is unsatisfying (too
sharp).

In order to get better results we had to suppose that the narrow
components are produced in a nearly pole-on \textcolor{black}{optically
thin} disk. We got satisfying results taking an inclination to the
sky plane of $12^{°}$, and as a first step a keplerian rotational
velocity taken to be 340 $km\, s^{-1}$at the star surface. With such
an inclination and velocity we are in agreement with the
 $v\,\sin i=70\, km\, s^{-1}$ determination of Israelian et al. (\cite{Ia96}).
 Such a region of line formation, would, when seen nearly pole on,
 resemble Kastner's static model.

To get the broad line components we had to suppose the existence of
a flow falling on the disk (or alternatively escaping from the disk),
whose material is incorporated into (or alternatively ejected from)
the disk material at the Keplerian velocity in each point.

\begin{figure}[h]
\includegraphics[%
  bb=40bp 100bp 540bp 790bp,
  clip,
  width=7cm,
  height=9cm,
  angle=270]{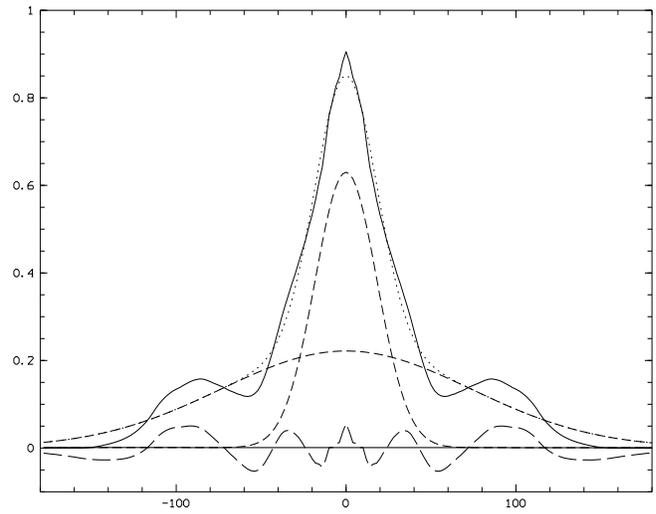}

\caption{Theoretical profile corresponding to the histogram of the emitting
material radial velocities of an optically thin Keplerian disk inclined
12° on the sky plane. An accreting wind generates the broad component.
\textbf{T}he total fit profile coincides with the broad component
for large velocities, and with the sum of the components for small
velocities. \textbf{}The component FWHM are the ones derived in Sec.
 \ref{sub:Line-components}. The Keplerian velocity is equal
to 340 $\mathrm{km\, s^{-1}}$ at the photospheric radius r=1. The
emission comes from $5<r<9$, in photospheric radius unit.
\label{Permitted_line_profile}}
\end{figure}

\begin{figure}[h]
\includegraphics[%
  bb=40bp 100bp 540bp 790bp,
  clip,
  width=7cm,
  height=9cm,
  angle=270]{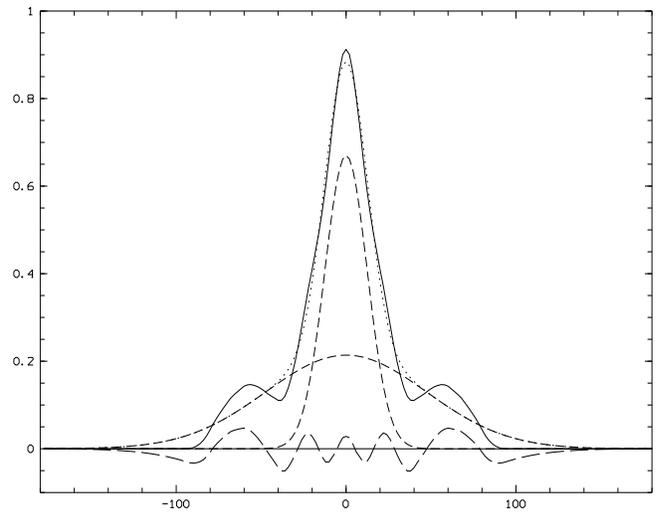}

\caption{Theoretical profile corresponding to the histogram of the emitting
material radial velocities of an optically thin Keplerian disk inclined
12° on the sky plane. An accreting wind generates the broad component.
The component FWHM are the ones derived in Sec. \ref{sub:Line-components}.
The Keplerian velocity is equal to 340 $\mathrm{km\, s^{-1}}$ at
the photospheric radius r=1. The emission comes from $15<r<20$, in
photospheric radius unit.\label{Forbidden_line_profile}}
\end{figure}

With such a configuration we can reproduce at the same time as well
the permitted (Fig.~\ref{Permitted_line_profile}) as the forbidden
(Fig.~\ref{Forbidden_line_profile}) Fe {\scriptsize II} line profiles,
by just playing on the distance to the star of the emitting zone.

The permitted and forbidden FWHM line components can be understood
if we assume that the permitted lines are formed in the inner parts
of the disk, while the forbiden ones issue from the external parts
of the Keplerian disk.

Fig.~\ref{Permitted_line_profile} shows the profile corresponding
to an optically thin Keplerian disk inclined 12° on the sky plane.
The velocity at the photospheric radius r=1 is equal to 340~$\mathrm{km\, s^{-1}}$.
The emission comes from the inner part ($5<r<9$, in photospheric
radius unit) of the optically thin disk whose total opening angle
is 24°. The components flux ratio narrow/broad is close to 1 (the
narrow component is slightly more intense) as was found for HD~45677,
while the components FWHM are similar (44 and 194~$\mathrm{km\, s^{-1}}$)
to the Fe {\scriptsize II} permitted lines ones.

Fig.~\ref{Forbidden_line_profile} shows the profile corresponding
to the same disk, but with emission coming from outer regions \\
($15<r<20$, in photospheric radius unit). The component flux ratio
narrow/broad is slightly higher than in Fig.~\ref{Permitted_line_profile},
as it is the case for {[}Fe{\scriptsize ~II}{]} with regard to Fe
{\scriptsize II} lines in HD 45677, while the components FWHM are
similar (28 and 108 $\mathrm{km\, s^{-1}}$) to the {[}Fe{\scriptsize ~II}{]}
lines ones. 

These two profiles model fairly well the mean component parameters
we measured in HD 45677 as well for the permitted as for the forbidden
Fe {\scriptsize II} lines.

\section{Discussion\label{sec:Discussion}}

\subsection{Level Population \label{sub:Level-Population}}

\begin{figure}[h]
\includegraphics[%
  bb=40bp 90bp 560bp 780bp,
  clip,
  width=65mm,
  keepaspectratio,
  angle=270]{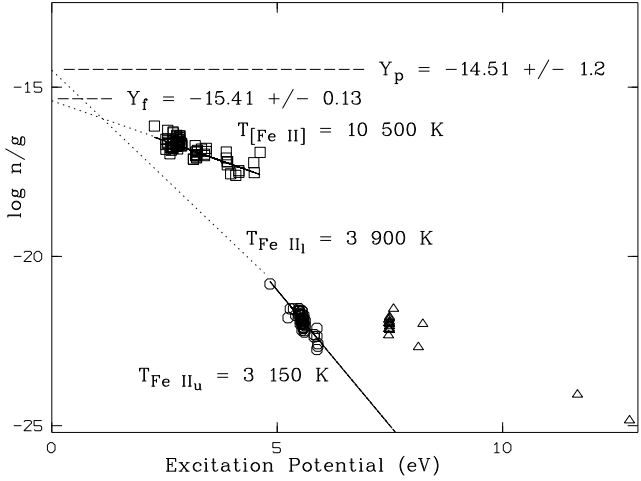}

\caption{Relative population of the $\mathrm{Fe^{+}}$ levels in the {[}$\mathrm{Fe\, II}${]}
(squares), and in the low and high excitation $\mathrm{Fe\, II}$
emission regions derived from the narrow emission components measured
in the 1999 spectrum of HD 45677. The fluxes of the lower excitation
emission lines (open circles) have been corrected for self--absorption,
as discussed in the text. The high excitation emission lines are drawn
as triangles. The straight lines represent the mean excitation temperatures
of the upper levels of the {[}Fe {\scriptsize II}{]} and of the lower
and upper levels of the Fe {\scriptsize II} emission lines, respectively.
The crossing at 4.72 eV marks the connection between the lower (3900
K) and upper (3150 K) Fe~{\scriptsize II} level population laws. The
zero points Y(0 eV) are the estimated ground level relative populations.\label{level_POP}}
\end{figure}

The results of our analysis concerning level population are summarised
in Fig. \ref{level_POP} where we plot for the three groups of lines
-- forbidden, low excitation and high excitation permitted Fe {\scriptsize II}
narrow components emission lines -- the logarithm of the normalised
line flux as a function of the excitation potential $\chi_{u}$ of
the upper level of the transitions for our 1999 data. The fluxes of
the intermediate excitation permitted lines are corrected for the
self--absorption as discussed above. The diagram is used to investigate
the formation region of the three line groups. One normally expects
the upper levels of the permitted lines to be de-populated with respect
to the metastable levels. If no de-population occurs, the population
laws for the upper levels and the lower levels should join. In fig.
\ref{level_POP}, the mean population laws for the upper (3~150~K)
and the lower (3~900~K) cross at the mean excitation potential value
of 4.72 eV, to take into account the overall deviation of the population
of the upper (odd) levels of the Fe~{\scriptsize II} transitions
with respect to the population of the lower (even) ones (see Section
\ref{Fe-II-permitted-emission-lines}). \textcolor{black}{The extrapolated
ordinate value at 0 eV of this law, given by equation \ref{Y0}, for
the Fe}\textcolor{black}{\scriptsize ~II} \textcolor{black}{lines
($-14.51\pm1.2$) turns out to be close to the value for the {[}Fe}\textcolor{black}{\scriptsize ~II{]}}
\textcolor{black}{lines ($-15.41\pm0.13$). This will be discussed
in Section} \textcolor{red}{}\textcolor{black}{\ref{ssection:Column-density}.}

We get in this diagram a limit to the position of the metastable levels
in the permitted line formation region. As forbidden and low excitation
Fe{\scriptsize ~II} transitions have the metastable levels in common,
we may note that the {[}Fe {\scriptsize II}{]} transition upper levels
in the Fe {\scriptsize II} permitted lines region appear to lie in
this diagram aproximately 2 to 3 dex under the location of the {[}Fe
{\scriptsize II}{]} observed line ones. Consequently {[}Fe {\scriptsize II}{]}
lines in the permitted line formation region cannot be observed.

Finally, the high excitation $\mathrm{Fe^{+}}$ levels are clearly
overpopulated with respect to the upper levels of the lower excitation
Fe~{\scriptsize II} emission lines. We attribute this overpopulation
to the same fluorescence mechanism acting in other peculiar stars,
such as KQ Pup, $\eta$ Car, and RR Tel (Muratorio et al. \cite{MVFBR92};
Viotti et al. \cite{VRB99}; Hartman \& Johansson \cite{HJ2000})
whose optical and UV spectrum shows prominent Fe {\scriptsize II}
lines arising from high excitation levels. Like in these cases, the
far--UV radiation is most probably an important process of population
of the $\mathrm{Fe^{+}}$ upper levels in the envelope of HD 45677,
while probably dielectronic recombinations are unimportant.

\subsection{Column density \label{ssection:Column-density}}

From the position of the bend in the SAC plots of Fe{\scriptsize ~II}
lines (Fig. \ref{SACN99_1.09_1.60} and Fig. \ref{SACW99_1.09_1.61}
for 1999 data) we are able to determine the value of $\mathrm{X_{c}}$
and hence through eq. \ref{X0} the values of the $\mathrm{Fe^{+}}$column
density of the Fe{\scriptsize ~II} permitted lines emitting regions
for each studied epoch.

To derive the $\mathrm{Fe\, II}$ column density $\frac{\mathrm{N}_{c}}{g_{c}}$
, defined by eq. (3), in the narrow component Fe {\scriptsize II}
permitted line region, we have to estimate the velocity broadening
of the opacity profile $v_{c}$ in eq.\ref{X0}. The narrow components
of the emission lines of HD 45677 should be formed in a region with
a small velocity gradient in the line of sight, as discussed in Section
\ref{ssec:Model}, it is supposed to be an equatorial disk seen nearly
pole on. If we are observing a disk in rotation, without being in
its plane, the appropriate $v_{c}$ to use, to which the line broadening
along a line of sight is linked, is twice the thermal broadening velocity
that has been derived from the {[}Fe{\scriptsize ~II}{]} emission
lines, as discussed in Section \ref{sub:Fe-II-forbidden}. The local
velocity dispersion for a temperature $\mathrm{T_{th}=}11\,500\,\mathrm{K}$
is given by $v_{th}=0.215\,10^{5}\sqrt{\mathrm{T_{th}/PA}}$ where
$\mathrm{PA}=55.85$ for $\mathrm{Fe^{+}}$(Baratta et al. \cite{Ba98}).
For the narrow components $v_{c}=6.2\, km\, s^{-1}$. As for the broad
components, the appropriate $v_{c}$ to use is given by our model,
as the mean velocity along a line of sight, $v_{c}$ is equal to $7\, km\, s^{-1}$.

\begin{table}[h]

\caption{logarithm of the minimum $\mathrm{Fe^{+}}$column density $\frac{N_{c}}{g_{c}}$
(for an excitation potential $\chi_{c}=2.75\, eV$) of both components
emission line region at the different epochs studied. The first line
display (N) the narrow component region values, while the second line
displays (B) the broad component region ones. \label{Tab_column_density}}

\begin{tabular}{|c|c|c|c|}
\hline 
year&
1995&
1999&
2001\tabularnewline
\hline
\hline 
 $\log\frac{N_{c}}{g_{c}}$ (N)&
$14.6\pm0.1$&
$14.9\pm0.1$&
$14.73\pm0.1$\tabularnewline
\hline 
 $\log\frac{N_{c}}{g_{c}}$ (B)&
$14.4\pm0.1$&
$14.5\pm0.1$&
$14.5\pm0.1$\tabularnewline
\hline
\end{tabular}
\end{table}

The different epochs and region $\log\frac{N_{c}}{g_{c}}$ values
at $\chi_{c}=2.75\, eV$ are given in Table \ref{Tab_column_density}. 
 If we extrapolate this column density to 0 eV using 1.28 as the exponent
of the power law of the lower terms population we obtain
 $\log\,\overline{N}(\mathrm{Fe^{+}})=46.5$ in 1999. 
 It is not certain that assumption (d), in sec. 3,
about the same Boltzmann population distribution for all metastable
levels is correct for the more excited of these levels (see Verner
et al. \cite{VGBJID02}). Therefore, if we are very conservative and
suppose that the mean excitation temperature down to 0 ev is 10~000~K,
typical of HII regions, as well as being of the order of what we find
here for the less excited forbidden \textbf{}{[}Fe {\scriptsize II}{]}
\textbf{}lines with lower levels at or near 0 ev, the column density
values might be smaller by a factor of about 2.1 dex.

The different narrow and broad component column densities in Table
\ref{Tab_column_density} are very similar. The eventual column density
variations with time are of the order of the precision.

As for forbidden lines, Fig. ~\ref{SAC_O_thin-[FeII]_F} shows a
marginally significant bending of the data points towards large optical
thicknesses,  as if the strongest {[}Fe~{\scriptsize II}{]} lines
were slightly self--absorbed. The curve bend, if real, would correspond
in this figure to a self absorption factor of 3 for strongest of the
observed forbidden lines, much less than what is observed for the
strongest permitted lines, as the optical depth range is lower for
forbidden lines. It would correspond to an $\mathrm{Fe}{}^{\textrm{+}}$
column density about 100 times larger than that above derived for
the permitted lines, which seems to us unrealistic. 

The vertical overlapping of the multiplets provides the zero point
of the ordinates: $Y_{f}=-15.41\pm0.13$ (see Table \ref{TAB-fit_F}).
This quantity is related to the total number $\overline{N}(\mathrm{Fe^{+}})$
of $\mathrm{Fe^{+}}$ ions in the {[}Fe {\scriptsize II}{]} emitting
volume through the relationship (see Baratta et al. 1998):

\begin{equation}
\log\,\overline{N}(\mathrm{Fe^{+}})/\mathrm{d}^{2}=Y_{f}+\log\,\mathrm{U}+16.977\label{N(Fe)}\end{equation}
, where U is the partition function of $\mathrm{Fe^{+}}$.

Assuming for HD 45677 a distance of 420 pc (Zorec \cite{Zo98}), we
obtain $\log\,\overline{N}(\mathrm{Fe^{+}})=45.6\pm0.13$ which gives
in 1999, for a cosmic abundance, a total hydrogen mass of $1.7\,10^{26}g$,
with a precision factor estimated as 0.13. 

\textcolor{black}{Notice that the $\log\overline{N}(\mathrm{Fe^{+}})=46.5\pm1.3$
value of the Fe}\textcolor{black}{\scriptsize ~II} \textcolor{black}{permitted
emitting region,} \textcolor{black}{obtained with a value of 1.28
for the lower level constant p of eq. (1), is close to the forbidden
Fe}\textcolor{black}{\scriptsize ~II} \textcolor{black}{region one.}

\subsection{Size of the emitting region \label{Size}}

From the fit results summarised in tables \ref{TAB-fit_N} and \ref{TAB-fit_W}
we can calculate, using equations \ref{X0} and \ref{Y0}, the apparent
size of both emitting line regions. 

\begin{table}[h]

\caption{Emission line region minimum sizes at the different epochs studied
(in cm). The first line display the narrow component region values (N),
while the second line displays the broad component region ones (B).
\label{Tab_Radius}}

\begin{tabular}{|c|c|c|c|}
\hline 
year&
1995&
1999&
2001\tabularnewline
\hline
\hline 
$\log\mathrm{R'}$ (N)&
$13.1\pm0.5$&
$12.9\pm0.3$&
$12.5\pm0.4$\tabularnewline
\hline 
$\log\mathrm{R'}$ (B)&
$12.8\pm0.8$&
$13.0\pm0.4$&
$12.6\pm0.5$\tabularnewline
\hline
\end{tabular}
\end{table}

For 1999 we obtain $\log\mathrm{R'=12.90\pm0.34}$ for the narrow
line region. The values of $\log\mathrm{R'}$ in Table.~\ref{Tab_Radius}
are typical radii of the line emitting region, equal to $\sqrt{\frac{S\cos i}{\pi}}$.
The radius of a B2 V star is according to Drilling and Landolt (2000
in `Astrophysical Quantities'', page 389)) near $4\,10^{11}$ cm.
The size of the region where the narrow Fe\, II
 emisison lines originate, probably a disk, turns out to be 20 (from
9 to 44) times that of the stellar radius. The model we presented
in Section \ref{ssec:Model} is consistent with these results as the
permitted line profile similar to the one observed is emitted from
a region whose surface is that of a disk of 7 stellar radii. The assumption
has been made in the $\mathrm{R'}$ determination that the population
laws for the upper levels and the lower levels join at 4.72 eV. Their
values obtained are therefore minima, corresponding to maximum w values.

The $\log\mathrm{R'}$ values obtained from 1995 and 2001 data in
Table \ref{Tab_Radius} remain close to the limits evaluated from
the 1999 data in Table \ref{Tab_Radius}.

As for the broad component emitting region apparent surface, the values
of $\log R'$ are very close, at least in 1999 to the narrow component
one. This result is consistent with the model of a disk seen nearly
pole-on, developped in Section \ref{ssec:Model}, as the regions in
which each component is emitted, being seen under the same opening
angle, are supposed to have similar apparent surfaces. 

As for \textcolor{black}{the narrow component} Cr{\scriptsize ~II}
emitting region\textcolor{black}{, we used} $v_{0}=6\, km\, s^{-1}$
\textcolor{black}{(twice the $10\,000\,\mathrm{K}$ thermal} broadening
\textcolor{black}{velocity) and the values of Table.~\ref{TAB-fit_Cr},
hence  a} $\mathrm{Cr\, II}$ column density $\frac{\mathrm{N}_{0}}{g_{0}}=15.98\pm0.6$.
\textcolor{black}{The $\log\mathrm{R'}$ value derived in 1999 ($12.2\pm0.8$)
is only slightly} smaller than that of the Fe{\scriptsize ~II} emitting
regions.

\section{Conclusions}

The SAC method has given consistent results. We have obtained the
properties of what is probably a disk of HD 45677, including limits
on a characteristic column density and a characteristic radius of
the Fe {\scriptsize II} emitting region. These estimates require reasonable
assumptions, one of which is a true Boltzmann distribution of the
populations of the lower even metastable levels and another that the
Boltzmann--like law of the populations of the odd upper levels of
strong optical lines are not more than what would be expected from
a smooth fit between these odd and even population laws. We have demonstrated
that the well known split emission line profile is in fact due to
absorption components like in the ultraviolet. We have demonstrated
that emission lines have narrow and broad components. We have shown
that the double emission profile can be explained with a disk plus
a flow (perhaps a wind) moving in or out of the disk. The Fe{\scriptsize ~II}
forbidden lines show a similar two component structure, but the smaller
width of each component is consistent with their production in a more
external part of the disk, where the Keplerian velocity is lower.
The examination of the fluxes of the narrow {[}Fe {\scriptsize II{]}}
components has shown that the metastable $\mathrm{Fe}{}^{\textrm{+}}$
levels follow a Boltzmann--type law with a mean excitation temperature
around 11~000~K. Such a value may actually represent the electron
temperature of the {[}Fe {\scriptsize II}{]} emitting region, if the
metastable levels are in equilibrium with the ground term through
electron collisions. This is probably the case of HD 45677, since
we expect the electron density to be large enough (N$_e \ge 10^6$ cm$^{-3}$)
for bringing the metastable levels to LTE (Nussbaumer \& Storey \cite{NS88}).
The low electron temperature with respect to the stellar surface temperature
may be due to cooling of the {[}Fe {\scriptsize II}{]} region by the
large line emission of this ion. Indeed, low electron temperatures
were predicted by Verner et al. (\cite{VVKFHF99}) and Verner et al.
(\cite{VGBJID02}) for the $\eta$ Car {\it Weigelt blobs} that are
strong $\mathrm{Fe}{}^{\textrm{+}}$ emitters. Fairly low {[}Fe {\scriptsize II}{]}
excitation temperatures were derived for $\eta$ Car (12000 K, Viotti
et al. (\cite{VRB99}) and for the symbiotic star RR Tel (6600 K Kotnik-Karuza
et al. \cite{KFS02}). 

The narrow components of the Fe {\scriptsize II} permitted emission
lines in HD 45677 are slightly, but significantly broader than the
corresponding {[}Fe {\scriptsize II}{]} components. From the SAC analysis
of the Fe {\scriptsize II} emission lines we have derived an $\mathrm{Fe}{}^{\textrm{+}}$
lower level excitation temperature much smaller than that derived
from the {[}Fe {\scriptsize II}{]} lines, namely around 3~900~K.
We take these results as evidence of different (but possibly partially
overlapping) emitting regions. The derived excitation temperatures
of the lower and upper levels of the permitted Fe {\scriptsize II}
lines are much smaller than that expected if $\mathrm{Fe}{}^{\textrm{+}}$
were in thermodynamical equilibrium with the local plasma. Verner
et al. (\cite{VGBJID02}) calculated the formation of singly ionized
iron in the Weigelt blobs of $\eta$ Car. According to them Fe$^{\textrm{+}}$
emission is the most important coolant in the blobs. Fe {\scriptsize II}
emission originates at electron temperatures $5\,000\, K\leq T_{e}\leq7\,500\, K$.
This can also be seen by comparing their figs, 8 and 9. We argue that
for HD 45677 the upper levels of the permitted low excitation transitions
are mostly populated by the diluted stellar UV radiation followed
by spontaneous decay, and that in the Fe {\scriptsize II}  region
the radiation density is high enough to make the radiation induced
transitions more important than electron collisions. Similar considerations
can be made for Fe{\scriptsize ~II} and {[}Fe{\scriptsize ~II}{]}
broad components, as their emitting region parameters appear fairly
close to the narrow components ones (see Table.~\ref{TAB-fit_N}
and Table.~\ref{TAB-fit_W} for Fe{\scriptsize ~II} lines and Table.~\ref{TAB-fit_F}
and Table.~\ref{TAB-fit_G} for {[}Fe{\scriptsize ~II}{]} lines).

The Fe {\scriptsize II} metastable level, as well as the lower level
(Table.~\ref{TAB-fit_N}) excitation temperatures are suspected to
increase with time. The high excitation anomalous multiplets may indicate
the presence of inhomogeneities.

The shape of the iron broad component is insufficiently known now
to decide about the structure of the non-disk component. Nevertheless
the similarity of the physical parameters derived for narrow and broad
components (T$_{\textrm{exc}}$, $\log\frac{N_{O}}{g_{0}}$) suggest
that both emitting regions, if not the same (different FWHM) are physically
linked.

Our identification of the broad components, and the suspected variability
of both emitting regions physical parameters should stimulate in future
new higher quality observations (especially in the blue) in order
to study more precisely the shape of the emission lines.

\begin{acknowledgements}
\textcolor{black}{We are grateful to R.F Viotti for very constructive
suggestions.} This work is partly based on contract of the Consiglio
Nazionale delle Ricerche (CNR).
\end{acknowledgements}

\end{document}